\documentclass[]{aa}
\usepackage{natbib}

\def\Msun{$M_{\odot}$}
\def\Rsun{$R_{\odot}$}
\def\SB9{$S\!_{B^9}$}
\def\kms{km~s$^{-1}$}

\bibpunct{(}{)}{;}{a}{}{,}

\usepackage[dvips]{graphicx}
\usepackage{times}
\newcommand\ignore[1]{} 
\usepackage{amsmath}

\begin{document}
\title{Spectroscopic binaries among  Hipparcos M giants\thanks{Based
    on observations carried out at the Swiss telescope installed at
    the {\it Observatoire de Haute Provence} (OHP, France), and at the
    1.93-m OHP telescope}$^,$\thanks{Tables 2, 3, and 6 are only available in electronic form
at the CDS via anonymous ftp to cdsarc.u-strasbg.fr (130.79.128.5)
or via http://cdsweb.u-strasbg.fr/cgi-bin/qcat?J/A+A/}}
\subtitle{I. Data, orbits, and intrinsic variations}
\titlerunning{Spectroscopic binaries among M Giants I. Data}
\author{B. Famaey\thanks{Postdoctoral Researcher, F.N.R.S., Belgium}\fnmsep\inst{1}
\and D.~Pourbaix\thanks{Research Associate, F.N.R.S., Belgium}\fnmsep\inst{1}
\and A.~Frankowski\thanks{Postdoctoral Researcher, F.N.R.S., Belgium.
Currently at Department of Physics, Technion-Israel Institute of
Technology, Haifa 32000, Israel}\fnmsep\inst{1}
\and S. Van Eck$^{\dagger}$\fnmsep\inst{1}
\and M.~Mayor\inst{2}
\and S.~Udry\inst{2}
\and A. Jorissen\inst{1}
}
\institute{
Institut d'Astronomie et d'Astrophysique, Universit\'e
libre de Bruxelles, Facult\'e des Sciences, CP. 226, Boulevard du Triomphe, B-1050
Bruxelles, Belgium   
\and
Observatoire de Gen\`eve, Universit\'e de Gen\`eve, CH-1290 Sauverny, Switzerland
}

\date{Received date; accepted date}

\abstract
{This paper is a follow-up on the vast effort to collect radial
  velocity data for stars belonging to the Hipparcos survey.} 
{We aim at extending the orbital data available for binaries with M giant primaries. The data presented in this paper will be used in the companion papers of this series to (i) derive the binary frequency among M giants and compare it to that of K giants (Paper~II) and (ii) analyse the eccentricity -- period diagram and the mass-function distribution (Paper~III).}
{Keplerian solutions are fitted to radial-velocity data. However, for several stars, no satisfactory solution could be found, even though the radial-velocity standard deviation is greater than the  instrumental error,  because M giants suffer from intrinsic radial-velocity variations
  due to pulsations. We show that these intrinsic radial-velocity variations can be linked with both the average spectral-line width and the photometric variability.}
{
We present an extensive collection of spectroscopic orbits for M giants with 12 new orbits, plus 17 from the literature. On top of these, 1 preliminary orbit yielded an
approximate value for the eccentricity and the orbital period. Moreover, to illustrate how the large radial-velocity jitter
  present in Mira and semi-regular variables may easily be confused
  with orbital variations, we also present
  examples of pseudo-orbital variations (in S UMa, X Cnc, and possibly
  in HD 115521, a former IAU radial-velocity standard). Because of
  this difficulty, M giants involving Mira variables were excluded
  from our monitored sample. We finally show that the majority of M giants detected as X-ray sources are actually binaries.}
{The data presented in this paper considerably increase the orbital
  data set for M giants, and will allow us to conduct a detailed analysis of the eccentricity -- period diagram in a companion paper (Paper III).}
\maketitle

\section{Introduction}
\label{Sect:Intro}

When a mass-losing M giant is present in a binary system, the interaction of the
wind of the giant with the companion gives rise to photometric
activity or spectroscopic symbiosis (like in symbiotic stars, in
VV~Cephei-like systems, or in low-mass X-ray binaries like V2116~Oph)
that make the system very conspicuous even far away.  However, if the
M giant does not lose large amounts of mass, the binary nature of the
star will not be as conspicuous. Combined with M giants often
exhibiting pulsations that cause intrinsic velocity variations,
thereby confusing the search for orbital variations, it explains why so few spectroscopic binaries involving M giants are known so far.  Indeed, the Ninth Catalogue of Spectroscopic Binary Orbits \citep[\SB9;][]{Pourbaix-04a}  contains 2746 entries (query in March 2007), among which only 32 systems involve M giants, and yet 21 of these are either well-known symbiotic systems or VV-Cephei-like systems.

Carquillat and collaborators have devoted a series of papers to spectroscopic
binaries of spectral types F to M, but only 2
binaries with primaries of spectral type MIII were studied
\citep{Carquillat-96,Prieur-2006}. Previously, \citet{Stephenson-1967} provided a list of 7 systems with composite spectra involving an M star. But it is the paper by
\citet{Hinkle-2002} that, to the best of our knowledge, has so far
provided 
the most extensive list of spectroscopic binaries involving non-symbiotic M giants. 

With the present paper, we start a series devoted to a detailed study of the properties of spectroscopic binaries involving an MIII primary. The number of such binaries with known orbital elements has nearly doubled, as a result  
of our observing campaign of an extensive sample of M giants, drawn
from the Hipparcos Catalogue, for which
CORAVEL radial velocities have been obtained in a systematic way
\citep{Udry-1997}.  
The main driver behind this large database lies, of course, with the stellar kinematics in our Galaxy. And indeed the
kinematical  
properties of the present sample of M giants have been fully analysed
by \citet{Famaey-2005} and \citet{Famaey-2008:a}.  
But this large data set may also be used to search for binaries. 

The present paper presents the radial-velocity data (Sect.~2) and
orbital elements (Sect.~4) of the newly-discovered spectroscopic
binaries, for which a satisfactory orbit could be obtained. It also
discusses the intrinsic variations sometimes mimicking orbital
variations (Sect.~3). The list of new orbital elements is complemented with an exhaustive list of orbital elements for non-symbiotic M giants drawn from the literature (Table~\ref{Tab:orbits}). In Paper~II \citep{Frankowski-2008}, we will use the observational information gathered in this paper to derive the frequency of spectroscopic binaries among M giants, and compare it with that of K giants. Paper~III \citep{Jorissen-2008} will then present an in-depth analysis of the eccentricity--period diagram for M giants. 

\section{Radial-velocity data}
\label{Sect:sample}

The basic sample of M giants
is drawn from the Hipparcos survey stars 
\citep[identified by flag 'S' in field H68 of the Hipparcos
catalogue;][]{ESA-1997}.
These M
giants were extracted from the Hipparcos survey stars on the basis of the
spectral type provided in the Hipparcos catalogue and of the absolute
magnitude $M_{H_p} < 4$ computed from the Hipparcos parallax and $Hp$
magnitude. Mira stars or supergiants of luminosity class I (when
explicitly mentioned in the spectral classification) were not included
in the sample (notably because of the confusion that their envelope
pulsation may cause on the radial-velocity variations). This first
sample (defined here as sample~I) contains 771
 M giants with declinations greater than $-15^\circ$ and corresponds
to the sample of northern M giants analysed in Famaey et al.~(2005), 
to which 65 M giants with declinations between 0 and $-15^\circ$ have been added.
Two radial-velocity measurements, spanning at least one year, have been obtained with the CORAVEL spectrovelocimeter \citep{Baranne-79} for all stars of sample~I,
as part of a monitoring programme targeting
all Hipparcos survey stars later than about F \citep{Udry-1997}.
Note that 22 objects of the sample of \citet{Famaey-2005} are not present in sample~I because they had  only one radial
velocity measurement, making them unsuitable for
binarity analysis. Famaey's sample was also screened for other irregularities, such as
wrong Hipparcos spectral type or mistaken identity\footnote{
The only mistake found in \citet{Famaey-2005} was the star HIP~26247 = RR~Cam = BD +72$^\circ$275, which was wrongly assigned radial velocity
measurements from the binary star J275 in the Hyades cluster. The
former star should therefore be discarded from the CORAVEL sample.
}.
Extrinsic S stars were removed from the sample as well.

\begin{figure}[]
  \includegraphics[width=\columnwidth]{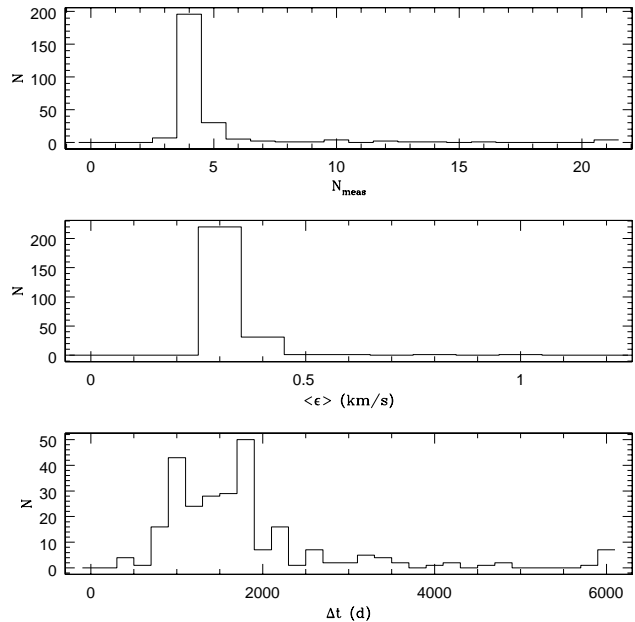}
\caption[]{\label{Fig:histoN}
Histograms of the number of measurements per star  ($N_{\rm meas}$),
the average uncertainty of one measurement ($<\!\epsilon\!>$), and the time spanned by the measurements of a given star ($\Delta
t$), for sample~II (see Tables~\ref{Tab:sample} and \ref{Tab:sampleII}).
}
\end{figure}

Subsamples of sample~I have been subject to more intensive observing campaigns. Every third star from this sample has received a denser coverage,
with (mostly) 4 instead of 2 measurements (7 stars received only 3
measurements), to achieve a better
binary detection rate: this subsample of sample~I is defined as sample~II.
Figure~\ref{Fig:histoN} displays the histograms of $N_{\rm meas}$ (the number of measurements per star),
$<\!\epsilon\!>$ (average uncertainty of one measurement), and $\Delta t$
(time spanned by the measurements of a given star) for the 254 stars of sample~II. 

\begin{table*}
\caption{\label{Tab:sample}
The four observing campaigns defining the four (sub)samples  of Hipparcos M giants.}
\begin{tabular}{ccccc}
\hline
\hline
Sample         & I & II & III & IV\\
\hline\\
Selection      & 2 measurements in Famaey et al.~(2005) &
$\sim$1/3 of I & all $\sigma(V_r) > 1$~\kms\ from II &  $\sigma(V_r) <  1$~\kms\ from II \\               &   &    &   & $0^h \le \alpha \le 16.5^h$ \\  
$N_{\rm star}$ & 771 & 254 & 35 & 138\\
$N_{\rm meas}$ per star  & 2 & $\ge 4^a$ &   $\ge 5$ & $N$(II)+1\\
Telescope      & Swiss 1-m (OHP) & Swiss 1-m (OHP) & 1.93-m (OHP) & 1.93-m (OHP)\\
Spectrograph   & CORAVEL & CORAVEL & ELODIE & ELODIE \\
Time span      & 1991--1999 & 1993--1995 & 2000--2004 & 2001--2004\\
\hline
\hline
\end{tabular}

$^a$ Seven stars have $N_{\rm meas} = 3$.
\end{table*}

Furthermore, 35 stars
from sample~II, suspected of being binaries (i.e., with
a radial-velocity standard deviation $\sigma(V_r) > 1$~\kms\ at the end of the observing campaign of sample~II)
have been monitored with the ELODIE spectrograph \citep{Baranne-96} at the Haute-Provence Observatory to derive their orbital elements. They make up sample~III. 
At the end of the ELODIE monitoring, a late measurement was obtained
for 157 stars from sample~II with  $\sigma(V_r) < 1$~\kms\ (this
selection being mainly based on right ascension), in order to detect binaries with very long orbital periods. These constitute sample~IV. A summary of the properties of these four samples is presented in Table~\ref{Tab:sample}.

The CORAVEL data were put on the ELODIE radial-velocity
system to ensure homogeneity \citep{Udry-1999}. The
uncertainty $<\!\epsilon\!>$ of one radial-velocity measurement is
approximately 0.3~\kms\ for CORAVEL measurements, but is better for
ELODIE measurements. With the ELODIE single-fibre mode used during the
present observing campaigns, it may be as good as 50 m~s$^{-1}$
\citep{Baranne-96}. It was not measured in a systematic way; however,  to fix the ideas, an accuracy of  0.2~\kms\ has been associated with the ELODIE measurements  in the data files.

\begin{table}
\caption[]{\label{Tab:sampleII}
The first five lines of the list of stars from sample~II. The full
table is only available electronically at the CDS. 
}

{\small              
\tabcolsep 2pt
\begin{tabular}{lccr@{.}lcccc}
\hline
\hline
HD    & $N_{\rm tot}$ & $N_{\rm ELO}$ & \multicolumn{2}{c}{$V_r$} & $\sigma(Vr)$ & $\epsilon(Vr)$ & $Sb^a$ & flag$^b$\\
       &     &  & \multicolumn{2}{c}{(km s$^{-1}$)} & (km s$^{-1}$) & (km s$^{-1}$) &(km s$^{-1}$) \\
\hline\\
1013  &5   &1  &$-$46&67&   0.18&  0.26&  2.80& -\\
1255  &5   &1  &   14&57&   0.32&  0.27&  3.95& - \\                               
2313  &5   &1  &   24&15&   0.18&  0.28&  3.15& -\\                         
2411  &6   &2  &    4&67&   1.06&  0.25&  4.58& ORB       \\
2637  &5   &1  &   11&25&   0.19&  0.25&  1.33& -\\
...\\
\hline                                
\end{tabular}

$^a$ The average sigma of a Gaussian fitted
to the cross-correlation profile, corrected for the instrumental width
(7~km~s$^{-1}$ for CORAVEL at the Haute-Provence Observatory);\\
$^b$ The flag in this column is the same as in Table~4.  
}
\end{table}

The data for sample~I (average velocity, radial-velocity standard
deviation, and binarity flag) may be found in Table~A.1\footnote{As a
  result of the confusion between HIP~26247 and
  a binary star (see footnote 1), the star
  HIP~26247 is erroneously flagged as a binary in that
  table.} (columns 24--26) of \citet{Famaey-2005}.  
A similar table, merging CORAVEL and ELODIE data, is given in
Table~\ref{Tab:sampleII} for sample~II, which also provides the binarity diagnostics (according to the rules that will be specified in
Sect.~\ref{Sect:diagnostics}).
Individual radial-velocity measurements for the 35 stars of sample~III
are given in Table~\ref{Tab:data}. Roger Griffin has kindly provided us with supplementary measurements for HD~182190 and HD~220088 that allowed us to compute an orbit for these two stars. These measurements were performed with his Cambridge spectrovelocimeter and are listed in Table~\ref{Tab:data}, with the label `CAM'. To put them on the ELODIE system, an offset of $-0.8$~\kms\ has been applied. 

\begin{table}
\caption[]{\label{Tab:data}
The individual radial-velocity measurements for the 35 stars of sample~III. The         
full table is available electronically from the CDS, Strasbourg.
}
\begin{tabular}{lllll}
\hline
\hline
HD & JD$-$2\ts400\ts000 & $Vr$ & $\epsilon(Vr)$ & Inst$^a$\\
\hline\\
2411 &  48876.599  &  4.97& 0.32&  COR\\                                              
2411 &  49223.556  &  6.05& 0.33&  COR\\                                              
2411 &  49668.360  &  6.42& 0.31&  COR\\                                              
2411 &  50717.506  &  4.74& 0.32&  COR\\                                              
2411 &  52304.297  &  3.56& 0.20&  ELO\\                                              
2411 &  53047.246  &  4.39& 0.20&  ELO\\                                              
4301 &  48441.832  &  6.36& 0.28&  COR\\                                              
4301 &  49309.361  &  7.08& 0.29&  COR\\                                              
4301 &  49581.863  &  6.96& 0.27&  COR\\                                              
4301 &  49602.549  &  6.58& 0.29&  COR\\                                              
4301 &  53048.254  &  4.50& 0.20&  ELO\\
...\\
\hline
\end{tabular}

$^a$ A flag identifying the spectrograph: COR = CORAVEL; ELO = ELODIE; CAM = Cambridge spectrovelocimeter
\end{table}

\section{Binaries, intrinsic radial-velocity 
jitter, and pseudo-orbits caused by pulsation}

\begin{figure}[]
  \includegraphics[width=\columnwidth]{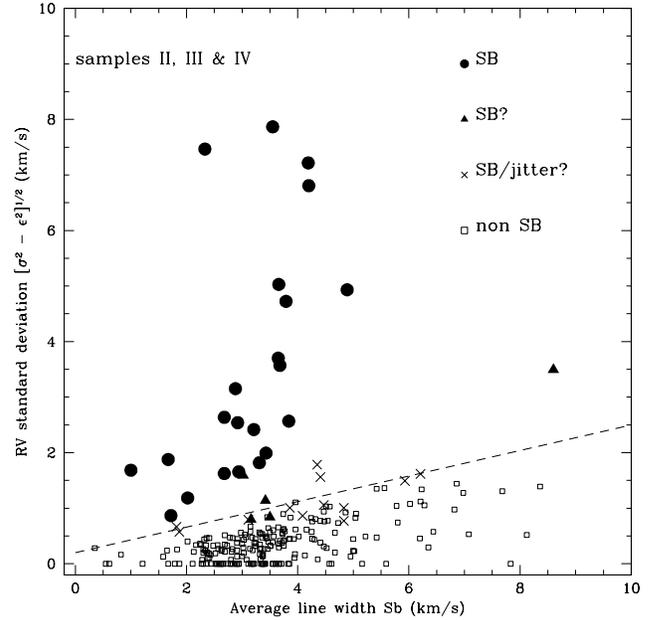}
  \caption{\label{Fig:Sb-sigma} 
The radial-velocity
standard deviation $\sigma_0(Vr)$  as a function of the average intrinsic line width
$Sb$, for samples~II, III, and IV (see Table~\ref{Tab:sample}). For non-binary stars (open squares), the maximum radial-velocity jitter increases with $Sb$.  The dashed line represents a rough division between  binary and non-binary stars. How this line has been defined will be discussed in Paper~II.
Symbols are as indicated in the figure.}
\end{figure}

\begin{figure}[]
  \includegraphics[width=\columnwidth]{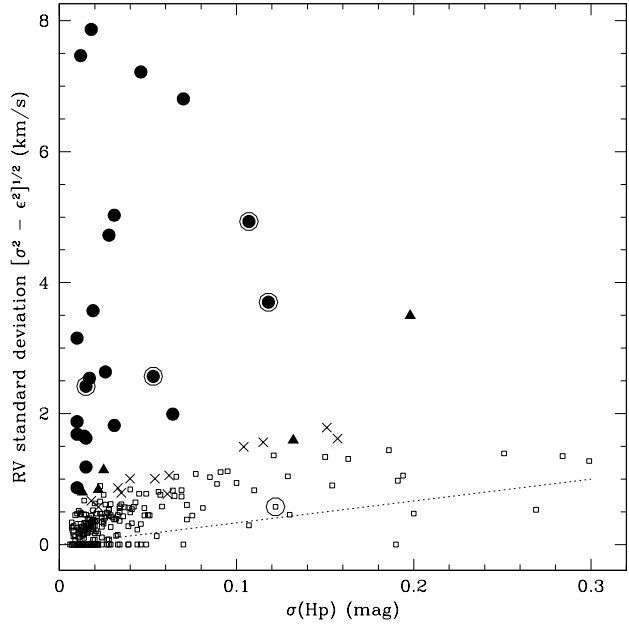}
  \caption{\label{Fig:variability} 
The radial-velocity standard deviation (corrected for
  the instrumental error) as a function of the standard deviation on the Hipparcos $Hp$ magnitude.
Symbols are the same as in Fig.~\ref{Fig:Sb-sigma}, except that large
circles identify stars detected as X-ray sources by the ROSAT
All-Sky Survey, following \citet{Hunsch98}.
The dotted line represents the trend between radial-velocity and
photometric variability reported  by \citet{Hinkle-1997}. This is {\it
  not} a dividing line like the one plotted in Fig.~2.} 
\end{figure}

\subsection{Binarity diagnostic}
\label{Sect:diagnostics}

The search for spectroscopic binaries (SBs) among M giants is made
difficult by the bulk mass motion existing in the atmospheres of these
stars \citep[all M giants being variable to some extent;
e.g. ][]{Eyer-1997,Jorissen-1997,Soszynski-2004}, since such motion triggers
some intrinsic radial-velocity jitter\footnote{A term introduced by \citet{Gunn-Griffin-1979} in this context.} \citep[e.g.,
][]{Udry-98a,Hinkle-2002}. This intrinsic jitter amounts to nearly
1.5~\kms\ in the coolest (non-Mira) M giants\footnote{\citet{Mayor-84} have shown that stars located at the tip of the giant branch in the globular
cluster 47~Tuc have a jitter of $\sim1.25$~\kms, while the jitter
reduces to 0.27~\kms\ for 1~mag-fainter stars. This trend goes further
down the red-giant branch, as shown with much higher accuracy levels
by \citet{Setiawan-2004} and \citet{daSilva-2006}.}
(Fig.~\ref{Fig:Sb-sigma}), so the detection of SBs cannot rely
solely on a $\chi^2$ test comparing the radial-velocity standard
deviation $\sigma(Vr)$ to the average instrumental error
$<\!\epsilon\!>$.
In the following, we denote by $\sigma_0(V_r)$ the radial-velocity
standard deviation corrected for the average
  instrumental error $<\!\epsilon\!>$, i.e. $\sigma_0(V_r) =
  (\sigma^2(V_r) - {<\! \epsilon \!>^2})^{1/2}$.

In a first step, to flag a star as a binary in samples~II, III, and IV,
we therefore rely solely on the visual examination of the
radial-velocity variations, and whether or not it is possible to obtain a meaningful orbital solution. 
The binarity diagnostic is then complemented by two criteria, described
in detail in Sects.~\ref{Sect:Sb} and \ref{Sect:Hp}:
(i) the location of the star in a diagram  $\sigma_0(V_r) - Sb$ (Fig.~\ref{Fig:Sb-sigma}),
where $Sb$ denotes the intrinsic width of spectral lines
(Sect.~\ref{Sect:Sb}), and (ii) its
location in a diagram $\sigma_0(V_r) - \sigma(Hp)$ 
(Fig.~\ref{Fig:variability}),
where $\sigma(Hp)$ is the standard deviation of the Hipparcos
magnitude (Sect.~\ref{Sect:Hp}).
The final binarity diagnostic for all stars from sample~III (and for
stars from sample~IV,  which are SB or suspected SB) 
is listed in Table~\ref{Tab:diagnostics} according to the following categories:
\begin{itemize}
\item[a] ``ORB" : a satisfactory orbit could be computed (see Fig.~4);
\item[b] ``ORB:" : the orbit is preliminary, because the number of data points
  or the time coverage are not large enough, or two different solutions are possible (see Fig.~4);
\item[c] ``SB" : the star is a spectroscopic binary, but there are not enough
  data points to even compute a preliminary orbit (see Fig.~5);
\item[d] ``SB?" : the star is suspected to be a binary (see Fig.~6), even though it falls close to the dividing line in the $\sigma_0(Vr)$--$Sb$ diagram (see Sect.~\ref{Sect:Sb} and Fig.~\ref{Fig:Sb-sigma}). 
\item[e] ``SB / jitter?" : it is not clear whether the radial velocity variations are of intrinsic or extrinsic nature (see Fig.~7): the star falls close to the dividing line in the $\sigma_0(Vr)$--$Sb$ diagram (Fig.~\ref{Fig:Sb-sigma}).  
\end{itemize} 
The radial-velocity variations (of intrinsic nature) of non-SB stars of sample~III are plotted in Fig.~8. 

\subsubsection{Intrinsic width of spectral lines}
\label{Sect:Sb}

We note that the parameter $Sb$, measuring the intrinsic width 
of spectral lines, provides useful guidance in distinguishing radial-velocity 
jitter from orbital motion. 
This parameter is defined as $Sb = (s^2 - {s_0}^2)^{1/2}$, where $s$ is 
the standard deviation of a Gaussian fitted to the CORAVEL cross-correlation 
dip \citep[see ][]{Baranne-79} and $s_0$ the instrumental width  \citep[7~km~s$^{-1}$ for CORAVEL at the Haute-Provence Observatory;
for details see][ also Paper~II]{Jorissen-Mayor-88,VanEck-Jorissen-00}.  
Indeed, Fig.~\ref{Fig:Sb-sigma} reveals that stars for
which no orbital solution could be found (open squares) have
radial-velocity standard deviations that increase 
as $Sb$ increases. In Paper~II, we will show that $Sb$ correlates closely
with the stellar radius. We thus foresee that radial-velocity
variations in stars falling well below the dashed line in
Fig.~\ref{Fig:Sb-sigma} are most likely of intrinsic (rather than
orbital) origin. Of course, binaries with an orbit seen almost face
on, or with a long period, or insufficiently sampled, may also fall
below the dashed line. 
Let us stress that the slope of this dividing line in Fig.~\ref{Fig:Sb-sigma} can be influenced by both the number of radial-velocity data points for a given object and the total number of objects in the sample (see Sect.~2.1 of Paper~II).

Examples of radial-velocity variations for such stars from sample III, falling below the dividing line in Fig.~\ref{Fig:Sb-sigma}, are plotted in Fig.~\ref{Fig:jitter}, and
it is clear that no orbital solution can be found for them. On the
other hand, stars along the dashed line, whose velocity variations are
plotted in Fig.~\ref{Fig:SB/jitter}, could either be SB or exhibit
radial-velocity jitter.  These 10 stars are listed in
Table~\ref{Tab:diagnostics}c. 

 \begin{figure}
   \includegraphics[width=\columnwidth]{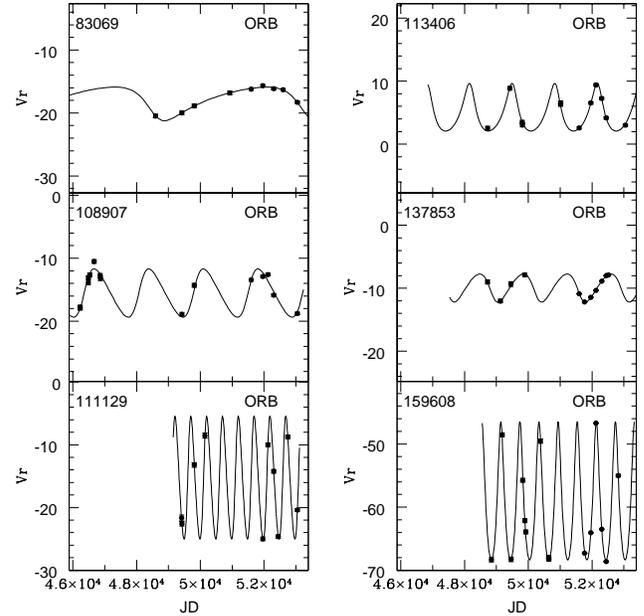}
\caption {\label{Fig:Vrbinaries} 
Radial-velocity data points for the spectroscopic binaries, along with their orbital solution.
 The solution marked as `ORB:' is not very well constrained and hence
 preliminary.
 All panels span a $Vr$ range of 30~\kms. Error bars on the individual measurements (of the order of 0.2~\kms) are too small to be seen.}
\end {figure}

\addtocounter{figure}{-1}
 \begin{figure}
   \includegraphics[width=\columnwidth]{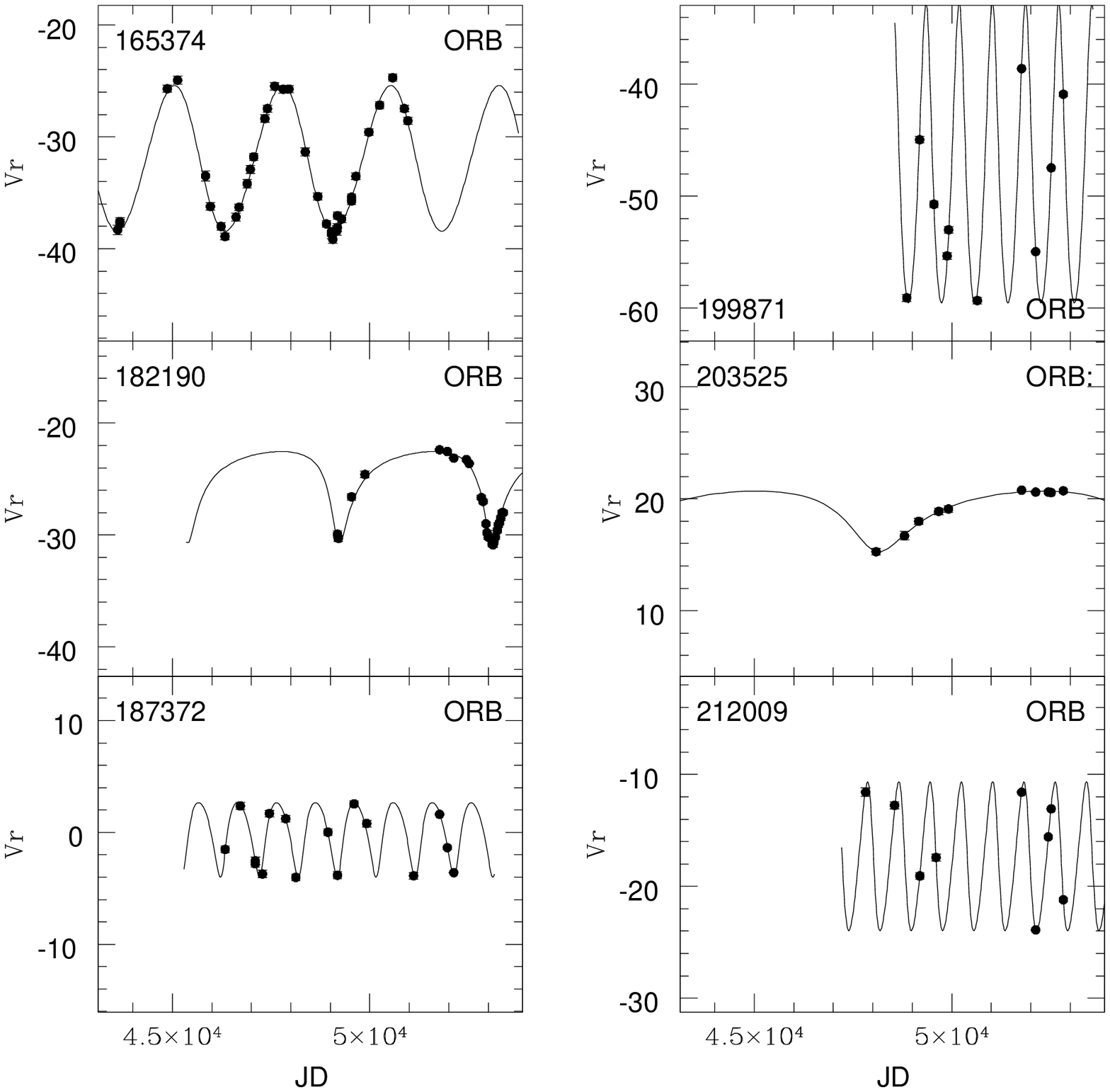}
   \includegraphics[width=\columnwidth]{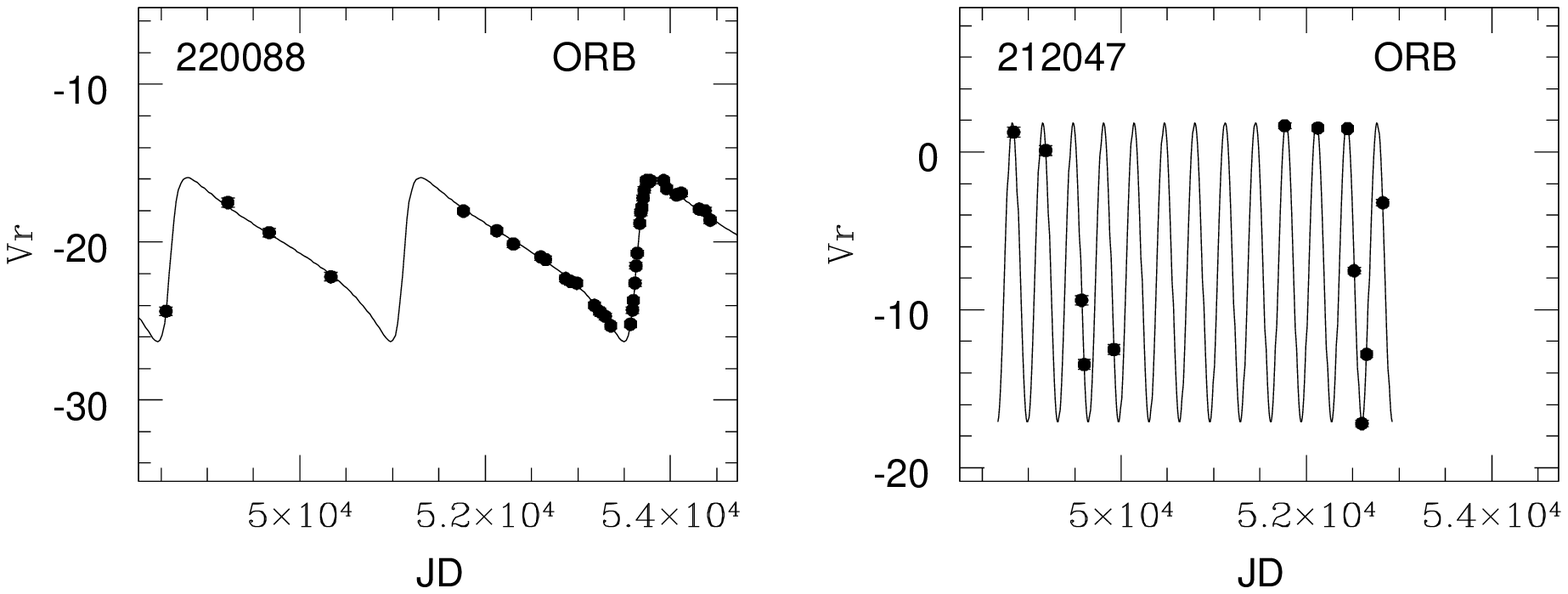}
 \vspace*{-6cm}
   \caption{Continued.}
 \end{figure}

\begin{figure}
  \includegraphics[width=\columnwidth]{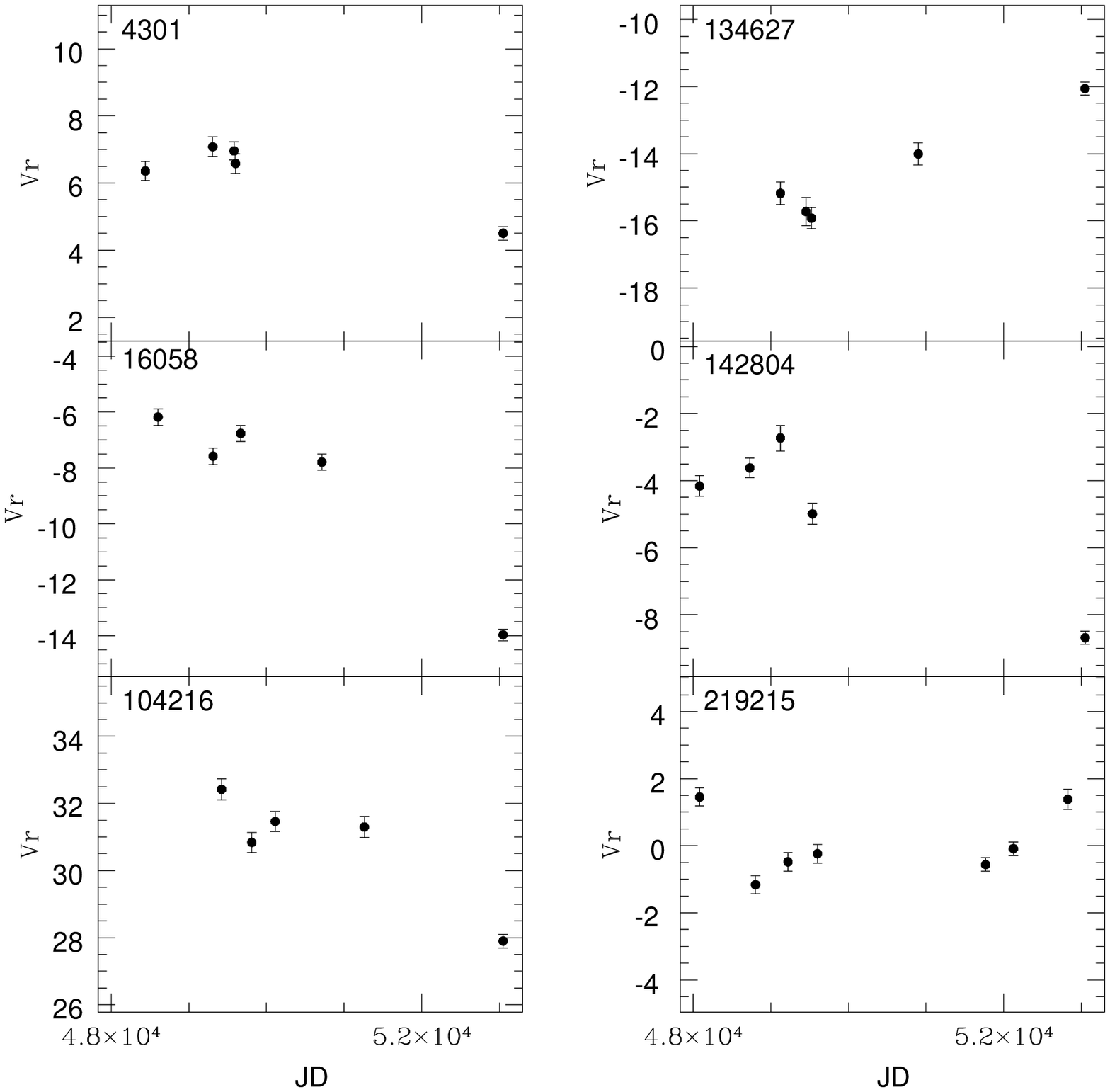}
  \caption{\label{Fig:SB}
Radial-velocity data points for the spectroscopic binaries with no
orbit available yet.}
\end{figure}

\begin{figure}
  \includegraphics[width=\columnwidth]{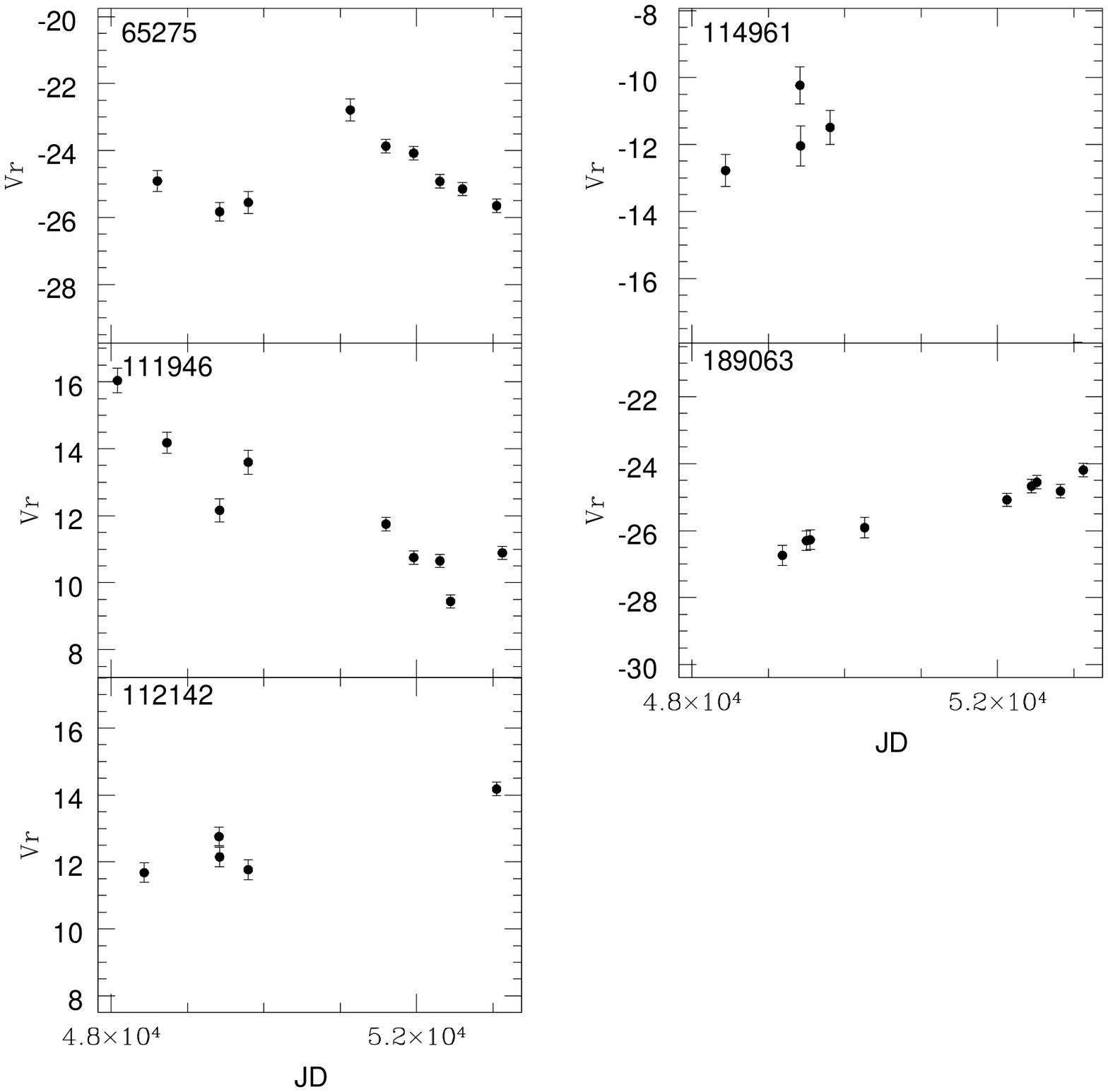}
  \caption{\label{Fig:SB?}
Radial-velocity data points for the suspected spectroscopic binaries.}
\end{figure}

\begin{figure}
  \includegraphics[width=\columnwidth]{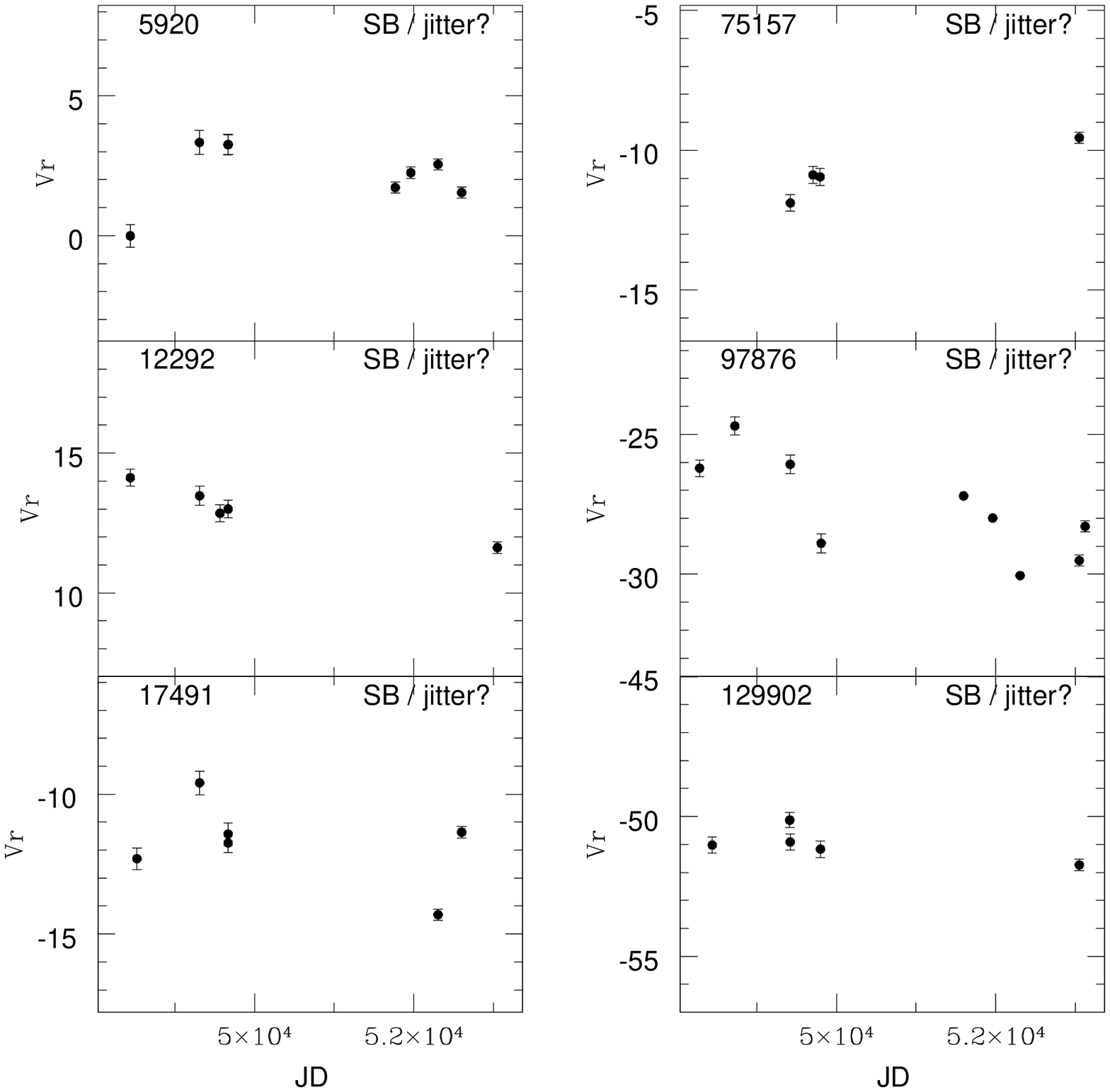}
   \includegraphics[width=\columnwidth]{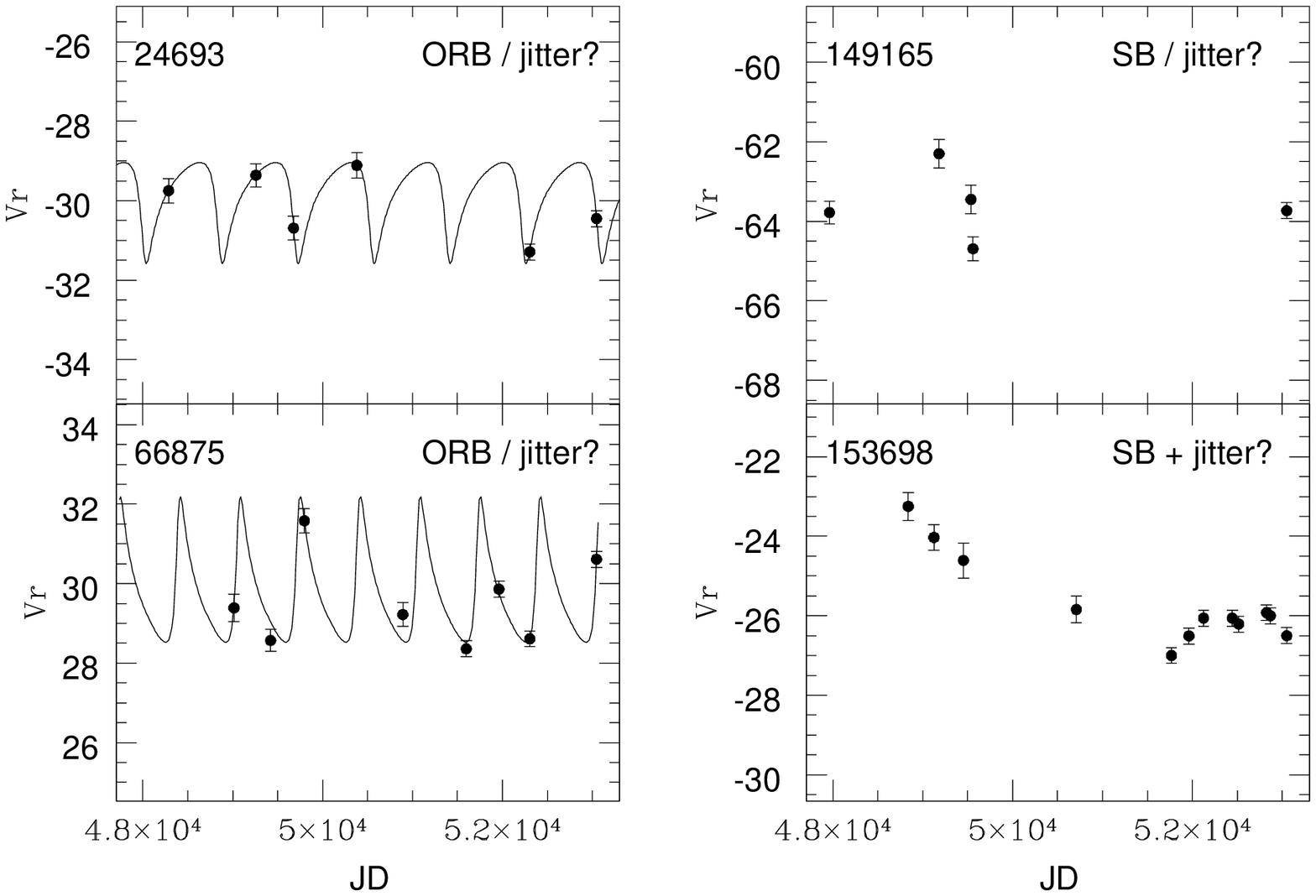}
\vspace*{-3cm}
  \caption{\label{Fig:SB/jitter}
Radial-velocity variations for stars falling close to the dividing line between jitter and orbital motion 
in Fig.~\ref{Fig:Sb-sigma}, for which the nature of the the radial-velocity variations is unclear. The vertical scale spans 10~\kms\ in all cases. They are denoted by crosses in Fig.~\ref{Fig:Sb-sigma}.
}
\end{figure}

\begin{figure}
   \includegraphics[width=\columnwidth]{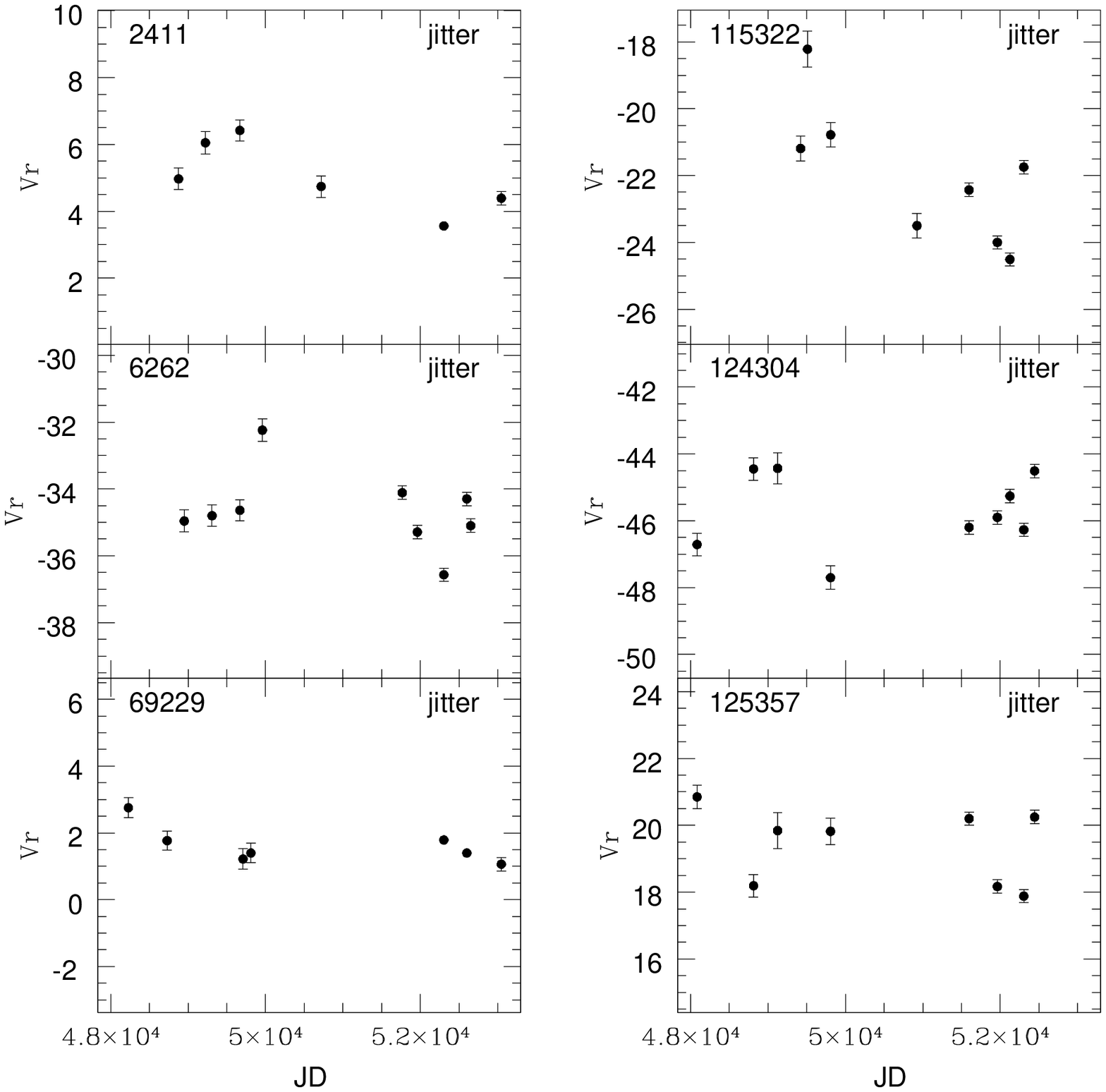}
   \includegraphics[width=\columnwidth]{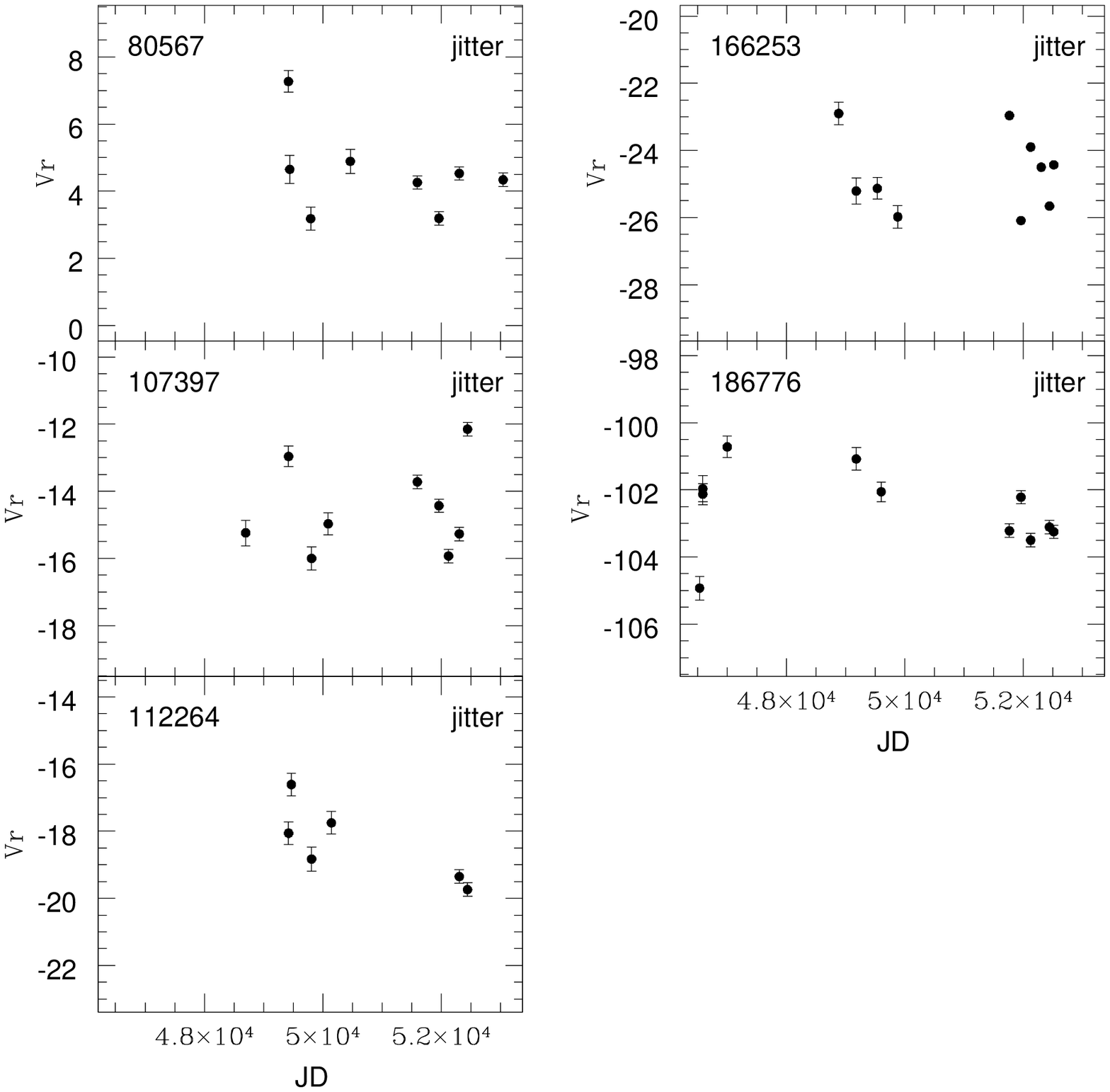}
  \caption{\label{Fig:jitter}
Examples of radial-velocity variations for  stars (referred to by their HD number, as listed in each panel) falling close to the dividing line between jitter and 
orbital motion 
in Fig.~\ref{Fig:Sb-sigma}, with a  radial-velocity jitter most
likely of intrinsic nature. They are denoted by open squares in
Figs.~\ref{Fig:Sb-sigma} and \ref{Fig:variability}.}
\end{figure}

On the other hand, among stars from samples~III and IV, for which 5 or more measurements have been made, 22 certain spectroscopic binaries have been found, as listed in Table~\ref{Tab:diagnostics}a. 
Of those, only 3 (HD~89758, HD~108907, and HD~132813) were previously known to be spectroscopic binaries, and a new satisfactory orbit could be computed for 12 of them (with the help of supplementary data points from the Cambridge spectrovelocimeter for HD~182190 and HD~220088; see Table~\ref{Tab:data}). The orbital elements are presented in Sect.~\ref{Sect:orbits}. The radial-velocity data points of the binary stars are displayed
in Fig.~\ref{Fig:Vrbinaries} for those binaries with an
orbit available (labelled  in Table~\ref{Tab:diagnostics} as ORB or,
for preliminary solutions, ORB:),
and in Fig.~\ref{Fig:SB} for the SBs without orbits.  
On top of the firm binaries, 5 stars were identified whose radial-velocity variations are very likely orbital (listed  in Table~\ref{Tab:diagnostics}b and displayed in Fig.~\ref{Fig:SB?}).


\begin{table*}
\caption{\label{Tab:diagnostics}
Binarity and photometric-variability diagnostics for stars from
sample~III (as well as binaries or suspected binaries from sample~IV).
}
\begin{tabular}{rrllccclcllll}
\hline\\
\noalign{{\bf a. Spectroscopic binaries} (marked as black circles in Figs.~2 and 3)}\\
\hline
\smallskip\\
HD  & $N$ &  $\Delta t$ &  Sp  &  $\sigma_0(Vr)$ &
\multicolumn{1}{c}{$<\!\epsilon\!>$}
 & $Sb$ & Flag & Rem. & GCVS$^a$ & Var. & $\sigma(Hp)$ & $P_{\rm phot}$\\
    &     &   (d)       &  & (\kms) & (\kms) & (\kms) & & & & & (mag) & (d)\\
\hline\\
83069 &    9 & 4450 &  M2III & 1.63 &0.22 & 2.68 &  ORB  &   & NSV 04545&    & 0.015\\
89758 &    7 & 6632 &  M0III & 3.57 &0.25 & 3.68 &  ORB  & $^b$ & NSV 04829& SR & 0.019\\
108907&   15 & 6821 &  M4III  &2.57 &0.25 & 3.84 &  ORB  &   & CQ Dra   & CV & 0.053 \\
111129&   10 & 3633 &  M2III  & 6.81&0.23 & 4.20 &  ORB  &   & BY CVn * &Irr & 0.070\\
113406&   12 & 4324 &  M1III  & 2.54&0.23 & 2.94 &  ORB  &   & *        &   & 0.017 \\  
132813&    5 & 3639 &  M4.5III& 4.93&0.27 & 4.89 &  ORB  & $^b$ & RR UMi   & SR& 0.107 & 43.3\\
137853&   11 & 3794 &  M1III  &1.66 &0.22 & 2.94 &  ORB  &   & *        &   & 0.014 \\  
159608&   16 & 3981 &  M2III & 7.87 &0.25 & 3.55 &  ORB  &   & *        &   & 0.018\\
165374&   37 & 7349 &  M2III &5.03  &0.33 & 3.61 &  ORB  &   & V980 Her *&Irr&0.031 \\
182190&   24 & 4197 &  M1III & 3.15 &0.22 & 2.88 &  ORB  &   & *        &   & 0.010\\
187372&   16 & 5787 &  M1III & 2.42 &0.26 & 3.21 &  ORB  &   & *        &   & 0.015\\ 
199871&   10 & 3964 &  M0III & 7.47 &0.24 & 2.33 &  ORB  &   &          &   & 0.012\\
203525&   10 & 4741 &  M0III & 1.68 &0.23 & 0.88 &  ORB :&   & *        &   & 0.010\\
212009&    9 & 5008 &  M0III & 4.72 &0.23 & 3.66 &  ORB  &   & KT Aqr * &   & 0.028\\ 
212047&   12 & 3984 &  M4III & 7.22 &0.22 & 4.19 &  ORB  &   & PT Aqr   & SR& 0.046\\ 
220088&   15 & 4743 &  M0III & 1.88 &0.22 & 1.67 &  ORB  &   & *        &   & 0.010\\
4301  &    5 & 4607 &  M0III &  1.18 & 0.25 & 2.02 & SB 
&   & * & & 0.015\\
16058 &    5 & 4443 &  M3III &  3.70 & 0.26 & 3.65 & SB 
&   & NSV 866 & Lb: & 0.118\\
104216&    5 & 3626 &  M2III &  1.99 & 0.27 & 3.43 & SB 
&   & FR Cam  & L   & 0.064\\ 
134627&    5 & 3922 &  M0     & 1.82 & 0.28 & 3.31 & SB 
&   & FF Boo * &    & 0.031\\
142804&    5 & 4964 &  M1III  & 2.64 & 0.27 & 2.68 & SB 
&   & NSV 7351& Irr & 0.026\\
219215&    7 & 4741 &  M2III  & 0.87 & 0.24 & 1.72 & SB 
&$^c$  & *        &    & 0.010\\ 
\hline
\end{tabular}

$^a$ An asterisk in this column means that  the variability 
  of the star has been discovered by Hipparcos.\\
$^b$ HD~89758, HD~132813: an orbit is also available in the literature, as listed in Table~\ref{Tab:orbits}.\\
$^c$ HD~219215: Preliminary orbital period: 2500~d\\

\end{table*}

\addtocounter{table}{-1}
\begin{table*}
\caption{(Continued). 
}
\begin{tabular}{rrllcclclllllll}
\hline\\
\noalign{{\bf b. Suspected binaries} (marked as black triangles in Figs.~2 and 3)}\\
\hline
\smallskip\\
HD  & $N$ &  $\Delta t$ &  Sp  &   $\sigma_0(Vr)$ 
 & $Sb$ & Flag & \multicolumn{2}{c}{Rem.$^a$} & GCVS & Var. &$\sigma(Hp)$& $P_{\rm phot}$\\
    &     &   (d)       &  & (\kms) & (\kms) & & & & && (mag) & (d)\\
\hline\\
65275&9& 4442 &  M0    & 0.84 & 3.50 &  SB?        & III&  &        &    & 0.022 \\  
111946&9&5040 &  M1/2III&1.59 & 3.02 & SB?         & III&  & VW Crv*& Irr & 0.132 & 45.6:\\
112142&5& 4613&  M3III & 1.14 & 3.42 & SB?        & IV &  & $\psi$ Vir&Lb & 0.025\\
114961&    5 & 4608 &  M7III & 3.50 & 8.60 & SB? & IV &  & SW Vir & SRb & 0.198 & 150 \\
189063&    8 & 3642 &  M0III  & 0.80 & 3.16 & SB? & III&$^d$  & *        &    & 0.013\\ 
\hline\\
\noalign{{\bf c. Binarity doubtful} (marked as crosses in Figs.~2 and 3)}
\smallskip\\
\hline
\smallskip\\
5920  &    8 & 4157 &  M4III &  0.77 & 4.83 &  SB/jitter? & III & & AK Cet & Lb & 0.061\\
12292 &    5 & 4606 &  M5III &  1.00 & 4.83 &  SB/jitter? & IV  & & AR Cet & SR:& 0.054\\
17491 &    6 & 4075 &  M5III &  1.62 & 6.21 &  SB/jitter? & IV  & & Z Eri  & SRb& 0.157 & 80.0\\
24693 &    6 & 4762 &  M1III & 0.79 & 3.11 &  ORB/jitter? & IV & $^b$ & GO Eri &    & 0.035\\
66875 &    8 & 4034 &  M3III & 1.06 & 4.47 &  ORB/jitter? & III& $^b$ & BL Cnc & L  & 0.062\\					    		
75157 &    4 & 3626 &  M0    & 1.01 & 3.86 & SB/jitter?   & IV &  & FW Cnc*& Irr& 0.040 \\
97876&9& 4841 &  M4III & 1.56 & 4.22 & ORB/jitter?        & III&  & UU Crt*& SR & 0.115  \\  		    	     
129902&    5 & 4605 &  M1III & 0.57 & 1.87 & SB/jitter? & IV & $^c$& *      &     & 0.022\\
149165&    5 & 5087 &  M1III & 0.67 & 1.82 & SB/jitter? & IV & $^c$& *      &     & 0.018\\
153698&   12 & 4209 &  M4III & 0.86 & 4.09 & SB + jitter?&III&  & *      &     & 0.033\\ 
\hline
\medskip\\
%
\noalign{{\bf d. Non-SB (radial-velocity jitter)} (marked as open squares in Figs.~2 and 3)}
\smallskip\\
\hline
2411&6 & 4171 &  M3III & 1.03 & 4.58 &&&& TV Psc     & SR & 0.085 & 49.1\\					    			     
6262  &    9 & 3702  & M3III & 1.04 & 4.51 &&&& V360 And * & SR & 0.066 &  \\
69229 & 7 & 4819 &  M1III & 0.44 & 2.32 &         & &  &        &    & 0.018\\
80567 &    8 & 3627  & M3    & 1.04 & 5.78 &&&& IN Hya     & SRb& 0.129 & 277.8    \\
107397&    9 & 3746  & M3III & 1.35 & 5.42 &&&& RY UMa     & SRb& 0.284 & 295.7   \\
112264&    6 & 3022  & M5III & 1.08 & 6.02 &&&& TU CVn     & SRb& 0.077 & 50.0\\
115322&    9 & 3627  & M6III & 1.49 & 5.93 &&&& FH Vir     & SR & 0.104 & 70.0   \\
124304&    9 & 4361  & M3III & 0.90 & 5.05 &&&& EV Vir     & SR & 0.154 & 120.0   \\
125357&    8 & 4361  & M2/3III & 1.78 & 4.35&&&&MZ Vir    & Irr& 0.151 &  \\
166253&   10 & 3631  & M0    & 1.11 & 3.96 &&&& V566 Her   & SRb& 0.091 & 137.0\\
186776&   11 & 5981  & M4III & 0.94 & 4.47 &&&& V973 Cyg   & SRb& 0.100 & 40.0\\
\hline
\end{tabular}

$^a$ This column specifies whether the
binarity suspicion resulted from campaign~III or IV (see
Table~\ref{Tab:sample})\\
$^b$ HD 24693 and HD 66875: Orbital solutions with an eccentricity of about 0.5 and periods of $845\pm7$ and $668\pm4$~d, respectively, are possible, but their reality is questionable. The stars are irregular (Lb) variables (varying from $Hp = 7.13$ to 7.26 and 5.94 to 6.11, respectively, in the
Hipparcos catalogue) falling right on the dividing line in Fig.~\ref{Fig:Sb-sigma}.\\
$^c$ HD~129902 and HD~149165 are probably spectroscopic binaries, given
their location in Fig.~\ref{Fig:Sb-sigma} just above the dashed
line. Unfortunately, the radial-velocity data presented in
Fig.~\ref{Fig:SB?} are too scarce to confirm that conclusion. \\
$^d$ HD~189063: Orbital period larger than 3600~d
\end{table*}
 
\subsubsection{Photometric variability}
\label{Sect:Hp}

Although Mira stars have been excluded from our sample, some
semi-regular variables are nevertheless present. Both classes
of variables often exhibit pseudo-orbital 
variations caused by
shock waves associated with the envelope pulsation \citep{Udry-98a,Hatzes-Cochran-1998,Wood-2000,Hinkle-2002,Setiawan-2004,Derekas-2006,Soszynski-2007,Hekker-2008}. 
For semi-regular variables,
\citet{Hinkle-2002} obtain semi-amplitudes $K$ between 1.6 and
3.1~\kms, whereas for Miras the semi-amplitudes may reach 20~\kms\ in
the most extreme cases \citep{Alvarez-2001}.  In terms of standard
deviations, these values become 1.1, 1.7, and 14~\kms, respectively
(remember that for sinusoidal variations, $\sigma = K /
\sqrt{2}$). Semi-regular and Mira variables may thus be expected to
exhibit $\sigma_0(Vr)$ values anywhere in the range of 1 to about
14~\kms. 

 Moreover, the radial-velocity curves of these stars may often be
fitted by a 
Keplerian orbit with a period of a few hundred days \citep{Udry-98a,Hinkle-2002,Wood-2004,Lebzelter-2005}.
However, it happens that this Keplerian solution  
fits the data for 4 to 10 cycles, and then becomes invalid. 
Two illuminating examples of this kind
of behaviour are discussed in the Appendix for stars that do not
belong to the samples considered 
in this paper (the Mira S star S~UMa and the semiregular carbon star X~Cnc). 
If the time span of the radial-velocity monitoring is not long enough
(i.e., shorter  than the 4 to 10 cycles mentioned above), the
inadequacy of the Keplerian solution may not be noticed, thus leading 
to the erroneous suspicion of binarity. 

It is therefore very important to check that the Keplerian
solutions proposed in Sect.~4 hereafter are valid, by eliminating
the possibility that they have an intrinsic origin, as often observed
for Mira and semi-regular variables. For this purpose, we collected
from the literature the photometric
properties of all the stars listed in
Table~\ref{Tab:diagnostics} (including those flagged as having
`radial-velocity jitter', among which we should expect to find a large
fraction of photometric variables). We searched the 
Hipparcos Variability Annex, which has the advantage of being an
unbiased information source containing all our stars. The results are plotted in
Fig.~\ref{Fig:variability} [$\sigma(Hp)$--$\sigma_0(Vr)$ diagram], and
listed in Table~\ref{Tab:diagnostics}. As we can see, all stars
flagged as spectroscopic binaries (Fig.~\ref{Fig:variability}), have
much larger radial-velocity standard deviations than expected from the relation between radial velocity and photometric variability for single stars (as reported by Hinkle et al. 1997). We also see that this relation exhibits some scatter (Fig.~\ref{Fig:variability}), and that stars flagged as `SB/jitter?' define a rough dividing curve between intrinsic and extrinsic radial velocity variations in the $\sigma(Hp)$--$\sigma_0(Vr)$ diagram.

\begin{figure}
  \includegraphics[width=\columnwidth]{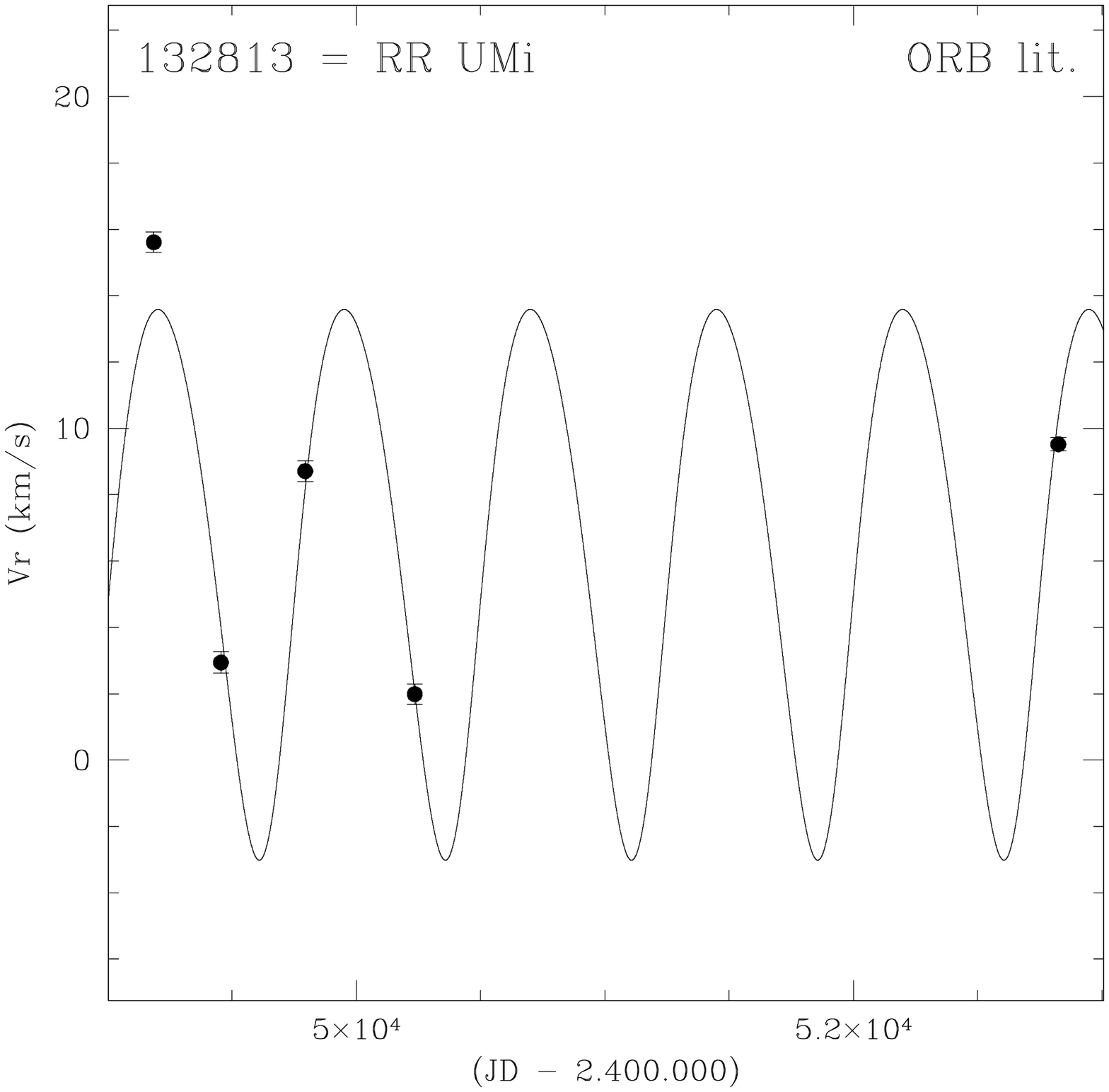}
  \caption{\label{Fig:HD132813}
Orbit from \citet{Batten-1986} for HD~132813. Overplotted are our new
data points, consistent with this orbit from the literature, obtained
about two decades ago.}
\end{figure}

In the column `Var' of Table~\ref{Tab:diagnostics}, we see that almost all stars later than M3 are semi-regular (SR) variables, as expected. It
is very clear that the fraction of SR variables with $\sigma(Hp) >
0.05$~mag increases among stars with (or suspected of having)
radial-velocity jitter. A specific case (HD~114961, marked as a suspected binary `SB?' in Table~\ref{Tab:diagnostics}) will be discussed in
Sect.~\ref{Sect:specific}. 
Such stars are not very numerous among the ones flagged as
SB. Therefore, one can be confident that all  
the stars flagged as SB in Table~\ref{Tab:diagnostics}a are indeed
binaries. Only two among those are semi-regular
variables  with large photometric amplitudes (i.e., HD~16058
= 15~Tri and HD~132813
= RR~UMi), and  there are
good arguments in favour of their binary nature. HD~16058 is a newly
flagged spectroscopic binary, but it is an X-ray source. As we show in
Sect.~\ref{Sect:X-ray}, X-ray detection in M giants is a strong
indication of binarity. The binary nature of HD~132813 has already been
claimed by \citet{Dettmar-1983} and \citet{Batten-1986}, who obtained consistent orbital
parameters (Table~\ref{Tab:orbits}). Our recent radial-velocity
measurements are consistent with these orbital parameters (see
Fig.~\ref{Fig:HD132813}), thus clearly demonstrating the binary nature
of HD~132813 from the stability of the Keplerian solution.

\subsection{X-ray emission as binarity diagnostic}
\label{Sect:X-ray}

X-rays are normally not observed in single
M giants lying to the right of the so-called dividing line, an almost
vertical boundary in the HR diagram separating stars with hot coronae
emitting X-rays (to the left) from stars with high mass loss (to the
right), which are normally not X-ray emitters \citep{Hunsch98}. Since
X-rays may be generated by several physical processes in a binary
system, their detection in an M giant is a strong argument in favour
of its binary nature. 
First, X-rays may be produced at the shocks resulting
from the collision of streams in the complex flow pattern associated
with wind accretion in a detached binary system involving an AGB star
\citep{Theuns-Jorissen-93,Theuns-96,Mastrodemos-98,Jahanara-2005}. 
Second, X-rays are generated when the
gravitational energy of the M giant wind falling in the potential well of
the companion star 
is converted into radiative energy when hitting the
stellar surface. The accretion luminosity
will either be radiated away in the form of hard X-rays if the
infalling
matter is optically thin; or if that matter is optically thick, half
will be converted into thermal energy and half will be radiated away
in the form of blackbody radiation (according to the virial theorem). 

A search for X-ray sources among M giants is therefore an
efficient way
of finding binary systems. 
The ROSAT all-sky
survey of X-ray sources detected only 11 out of 482 M giants of
luminosity classes I to III from the {\it Bright Star Catalogue}
\citep{Hunsch98}.  They are listed in Table~\ref{Tab:X}, along with a comment regarding 
the binary nature. These few detections justify using radial-velocity variations to detect binaries among M giants in our paper.

\begin{table}
\caption[]{\label{Tab:X}
Properties of the 11 M giants detected as  X-ray sources by the ROSAT
all-sky survey \citep[following][]{Hunsch98}, where for three stars, the X-ray source is offset by more than 1' from the M giant optical position.}

\begin{tabular}{rrlllllll}
\hline
HR & HD & Name & Sp. Typ. & Binary? \\ 
\hline
750 & 16058 & 15 Tri     &M3 III & SB\\ 
2216& 42995 &$\eta$  Gem &M3 III & ORB + visual GV comp.\\
3013& 62898 &$\pi$   Gem &M1 III & X-ray offset\\
4765& 108907& 4 Dra      &M4 III & ORB\\
5512& 130144& - &M5 IIIab & Vr constant?\\
5589& 132813& RR UMi &M4.5 III & ORB - X-ray offset \\
6200& 150450& 42 Her &M2.5 IIIab &\\
6374& 155035& - &M1-2 III & X-ray offset\\
6406& 156014&  $\alpha^1$ Her &M5 Ib-II & visual GV companion\\
7547& 187372& - &M1 III & ORB \\
8992& 222800& R Aqr &M7 IIIpe & SB (symbiotic)\\
\hline
\end{tabular}
\end{table}

It may be seen from Table~\ref{Tab:X} that, in most cases, the M
giants detected as X-ray sources are flagged as binaries (4 from the
present work, 1 from the literature, and 1 symbiotic). In three cases (HD~62898, HD~132813, and HD~155035),
the X-ray source is offset by more than 1' from the optical
position of the giant, which casts doubts on the M giant being the
source of the X-rays \citep[see ][ for details]{Hunsch98}. 
In two other  cases,  there is a
visual G-type companion, where X-rays may arise from
coronal emission. Thus only HD~130144 and HD~150450 have no
satisfactory explanation  for the
origin of the X-rays. In Fig.~\ref{Fig:variability}, it is HD~130144 that
appears amidst the non-binary M giants.

\subsection{Special cases}
\label{Sect:specific}
\subsubsection{HD~114961}

Since semi-regular and Mira variables may be expected to exhibit $\sigma_0(Vr)$ values anywhere in the range from 1 to about 14~\kms, the value $\sigma_0(Vr) = 3.1$~\kms\ observed for the
semi-regular variable HD~114961 (SW~Vir) in sample~II is not incompatible with
intrinsic variations (although we flagged it as a suspected binary; indeed, HD~114961 is the triangle in
Fig.~\ref{Fig:Sb-sigma} with the largest $\sigma_0(Vr)$ for $Sb =
8.5$~\kms).
Let us note, however, that the radial-velocity data available for this star are unfortunately too scarce to infer any periodicity, so no firm conclusion can be reached at this point regarding the intrinsic or orbital origin of the large radial-velocity scatter exhibited by SW Vir. 

\subsubsection{HD~115521}
\label{Sect:115521}

\begin{figure}[]
  \includegraphics[width=\columnwidth]{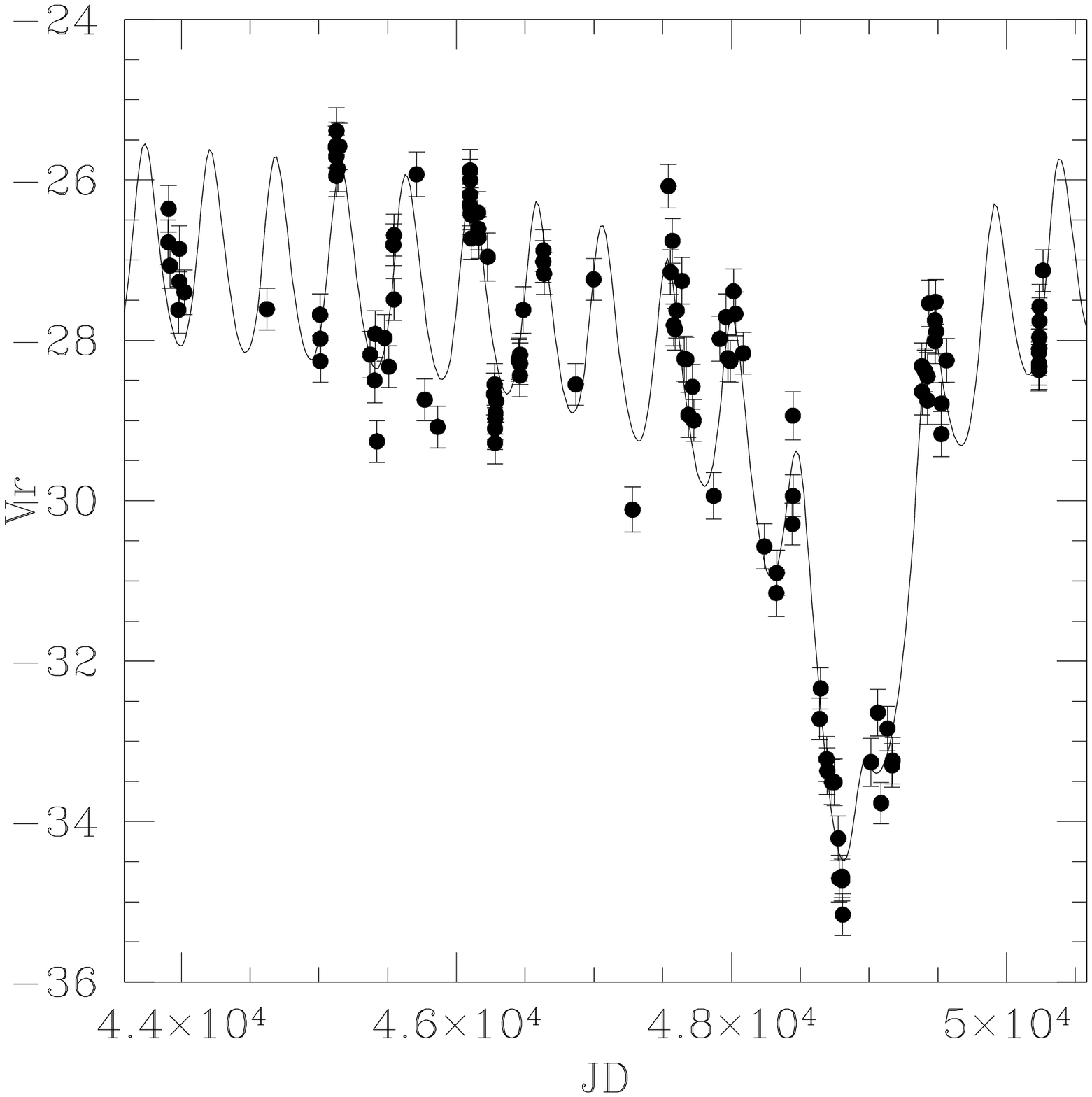}
  \caption{\label{Fig:115521}
  The radial-velocity measurements of the former radial-velocity standard HD~115521. Short-term variations with a period of 475~d are superimposed on variations with a much longer time scale. The solid curve is obtained by adopting $P = 29208$~d and $e = 0.87$ for the long-period orbit.}
\end{figure}

HD 115521 is an interesting case, similar to those discussed in the appendix. It belongs to sample~I (and therefore does not appear in
Table~\ref{Tab:diagnostics} since it has only been measured twice in our observing campaign), but being a radial-velocity
standard star till the 1990s \citep{Udry-1999b}, it was measured very
frequently and turned out in the first place to be variable with a small amplitude
 ($K \sim 1.5$~\kms) and a period around 500~d
\citep{Duquennoy-Mayor-91}. The 127 measurements are listed in Table~\ref{Tab:115521_data}, only available
electronically at the CDS, Strasbourg.
Later on, variations on a much longer time scale became apparent, exceeding the
measurements' time span of
6358~d. The orbital period therefore cannot be determined with good
accuracy. Moreover, owing to this insufficient  sampling of the orbital cycle, the value adopted for the orbital period strongly influences the value derived for the eccentricity. 
Table~\ref{Tab:115521} lists the {(pseudo-)orbital} elements used
to draw the radial-velocity curve of Fig.~\ref{Fig:115521}. The high
uncertainties on the elements of the long-period orbit 
should serve as a reminder that these
elements are very uncertain. Only lower bounds to the orbital period
and the eccentricity are therefore listed in
Table~\ref{Tab:orbits}.

The properties of the short-period variations would imply 
a rather low mass for the companion: assuming a mass of
1.3~\Msun\ for the giant, the minimum mass for the companion is
0.054~\Msun, corresponding to a brown dwarf.   
It is not entirely clear whether the short-period variations are
indeed due to an orbital motion, for several reasons. First, as for
S~UMa and X~Cnc discussed in the Appendix, the data deviate from the Keplerian
solution after a dozen cycles, and this is very clearly seen in
Fig.~\ref{Fig:115521},  where the first data points do not fall on the
solution defined by later measurements.  
Second, with $\log g = 0.70$
 and $K =
1.2$~\kms\ (Table~\ref{Tab:115521}), 
the short-period variations of HD~115521 fall on the $\log g
- K$ ($K$ being the semi-amplitude of the radial-velocity variations)
relationship as defined by \citet{Hekker-2008} for K giants\footnote{The gravity estimate is
  obtained as follows. From $V = 4.80$ and the 2MASS value $K = 0.47$,
one derives $T_{\rm eff} = 3680$~K and $BC_K = 2.74$, from the
$V-K$ calibrations of \citet{Bessell-Wood-84} and
\citet{Bessell-98}. The maximum-likelihood distance of
\citet{Famaey-2005} ($168\pm20$~pc) then yields $M_{\rm bol} = -2.9$,
and a radius of 84\Rsun\ from the Stefan-Boltzmann relationship
between luminosity, $T_{\rm eff}$ and radius. Adopting a mass of
1.3~\Msun\ then yields a gravity of $\log g = 0.70$.}. The
existence of this relation between the amplitude of the
radial-velocity variations and an intrinsic property of stars like the
surface gravity hints at an {\em intrinsic} origin of such
radial-velocity variations.  The relation  $\log g
- K$ found by \citet{Hekker-2008} among K giants \citep[initially
  suggested by ][]{Hatzes-Cochran-1998} continues in the domain of M
giants, as we discuss in Paper~II, where it is shown that the
radial-velocity standard deviation correlates with the CORAVEL
parameter  $Sb$ (measuring the spectral line width), which is in turn
a good measure of the stellar radius (see Figs.~1 and 3 of
Paper~II). 

However, if the shorter-period radial-velocity variations are intrinsic, then one expects photometric variations with a period of 475~d; unfortunately, we could not find any mention of them,
  in either the
  Catalogue of Suspected Variable Stars, where HD~115521 is entry NSV~6173,
  or in the Hipparcos Photometry Annex, or in the ASAS Catalogue of
  Variable Stars \citep[]{Pojmanski-2002}. 

\section{Orbital elements of newly-discovered binaries}
\label{Sect:orbits}

\begin{figure}[]
  \includegraphics[width=\columnwidth]{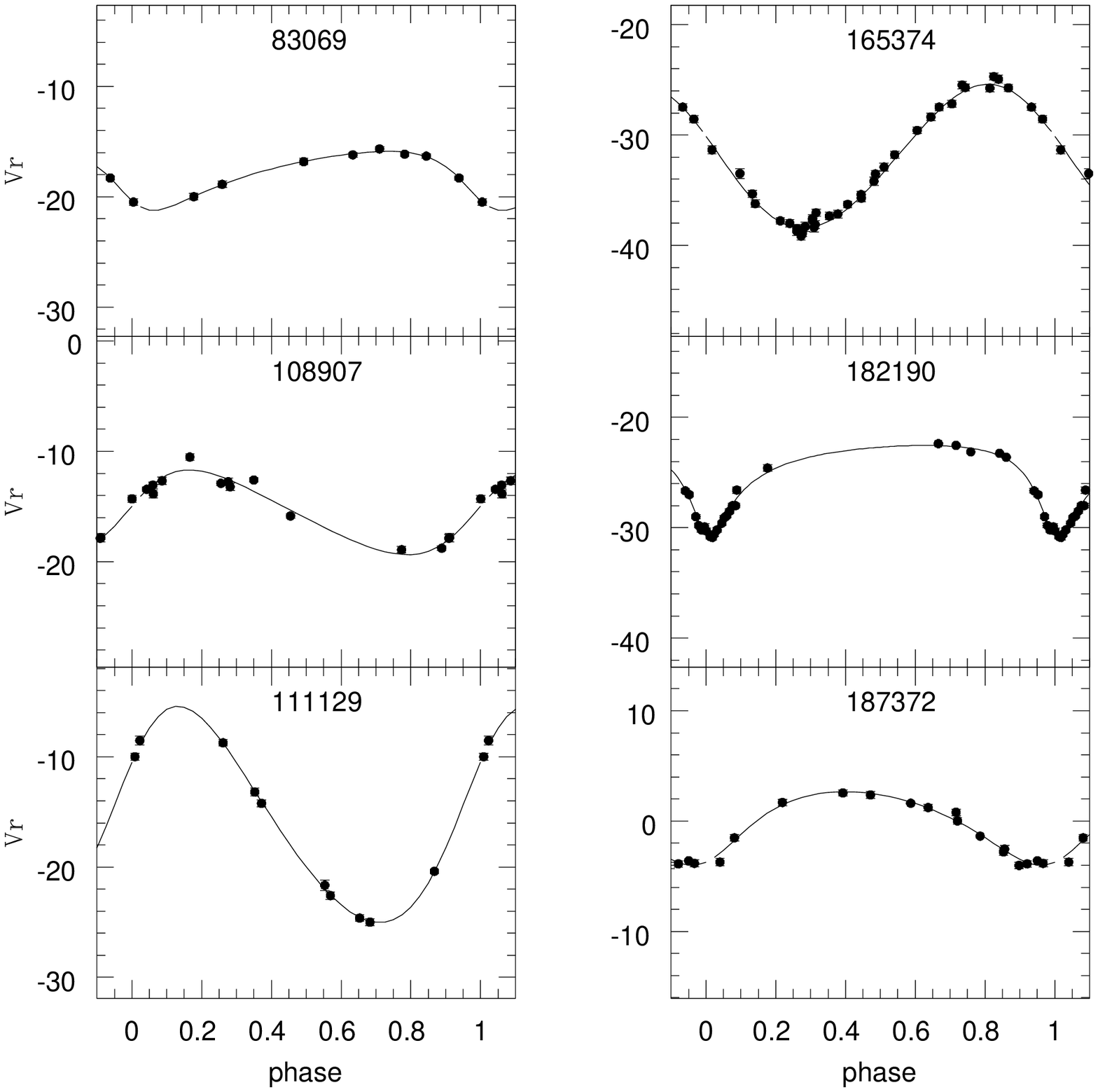}
  \includegraphics[width=\columnwidth]{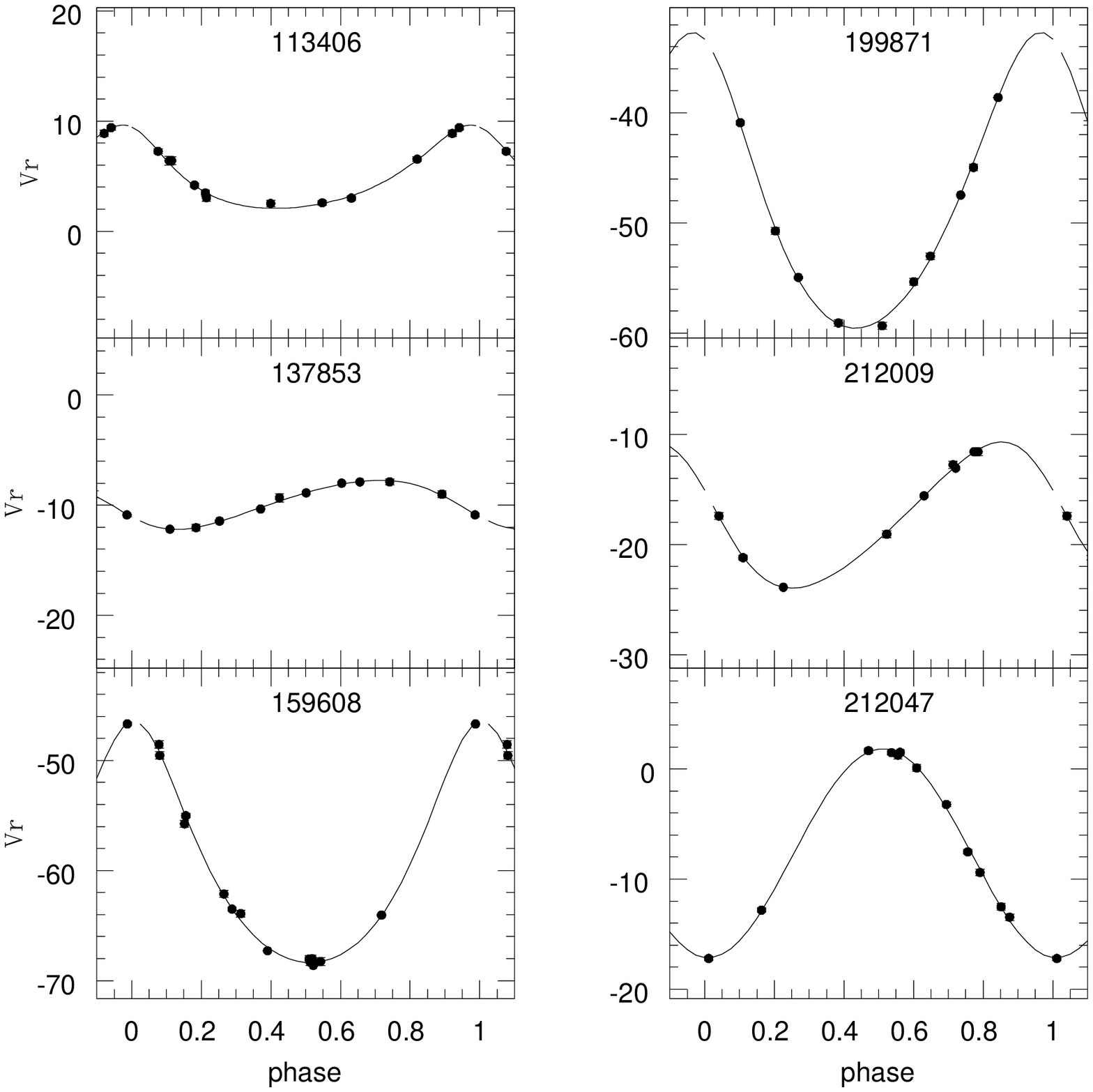}
   \includegraphics[width=\columnwidth]{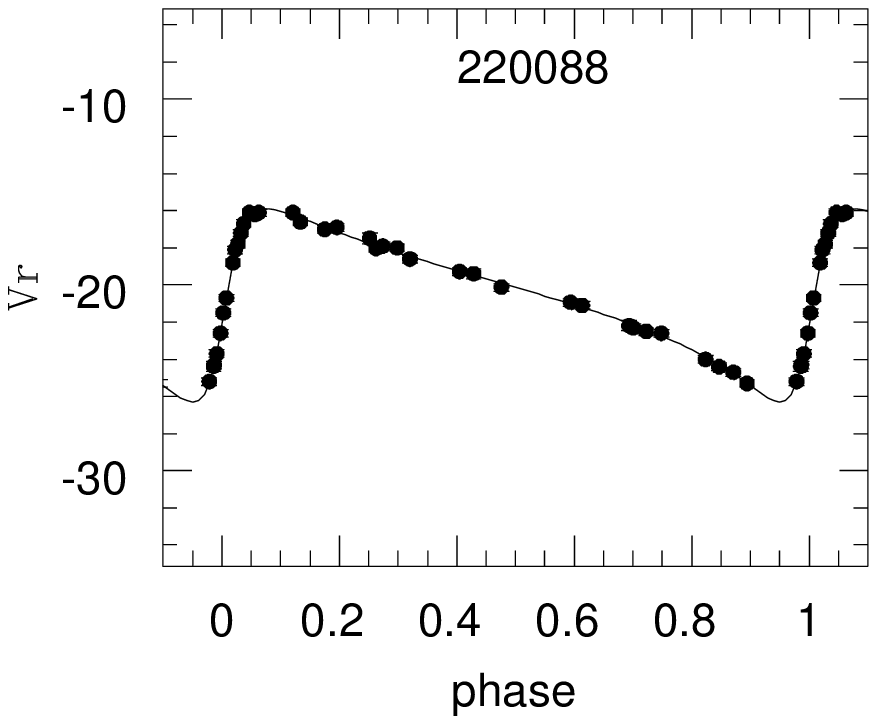}
\vspace*{-6cm} 
 \caption{\label{Fig:phase}
Orbital phase diagrams for the orbital solutions, according to the
solutions listed in Table~\ref{Tab:orbits}.
}
\end{figure}

The complete set of orbital elements for the 12 newly discovered
binaries are listed in Table~\ref{Tab:orbits}. Figure~\ref{Fig:phase}
presents the phase diagrams for those firm orbital solutions. The
second part of Table~\ref{Tab:orbits} provides a preliminary period
and eccentricity for the binary with not enough data points to derive
meaningful orbital solutions. For the sake of completeness, the last
part of Table~\ref{Tab:orbits} collects periods, eccentricities, and
mass functions for (non-symbiotic)  M giants available in the
literature, or kindly communicated by R. Griffin. 
This combined data set will be used in Paper~III to discuss general properties (like the eccentricity--period diagram) of systems
involving M giant primaries.
It must be stressed that Table~\ref{Tab:orbits} includes orbits  
neither for symbiotics nor for VV-Cephei-like systems (VV~Cep, AZ~Cas,
etc.). A
list of orbital elements for the former may be found in
\citet{Belczynski00} and \citet{Mikolajewska-03}.
\citet{Mikolajewska-2007} and \citet{Fekel-2007} provide references  
for the most recent symbiotic orbits.

\begin{table}
\caption[]{\label{Tab:115521_data}
The 127 radial-velocity measurements for HD 115521. The table, only available at the CDS, Strasbourg, lists the Julian Date, the radial velocity, and the corresponding error. 
}
\end{table}

\begin{table}
\caption[]{\label{Tab:115521}
Orbital elements for the system HD 115521. It is not certain, however,
that the short-period variations (A+a) are due to an orbital motion
(see text). The long period orbit is not well-constrained.  
}
\begin{tabular}{lllll}
\hline
         &   A+a        & Aa+b\\
\hline
$V_\gamma$ & $-26.93\pm2.1$ & 0 \\
$P$ (d) 	& $475.1\pm1.4$ & $29000:$\\
$T_0$ & $2\ts449\ts411\pm29$ & $2\ts419\ts767\pm86282$\\
$e$ & $0.148\pm0.062$ & $0.87\pm0.24$\\
$\omega$ ($^\circ$) & $-27.1\pm22.9$ & $191.8\pm3.5$\\
$K$ (\kms) & $1.25\pm0.07$ & $4.2\pm0.5$
\medskip\\
$f(M)$ (M$_\odot$) & $9.3\;10^{-5}$ & 0.027 \\
$a_1 \sin i$ (AU) & 0.053 & 5.57\\
\hline
\noalign{$\sigma(O-C) = 0.50$~\kms} 
\end{tabular}
\end{table}

\tabcolsep 3pt
\begin{table*}
\caption{\label{Tab:orbits}
An extensive list of orbital elements for (non-symbiotic) M giants. 
}
{\footnotesize
\tabcolsep 2 pt
\begin{tabular}{rrrcrlllclrrrll}
\noalign{\bf New orbits}\\
\hline
HD & Sp. & $N$ & $\sigma$(O$-$C) &
$\Delta t$ & 
\multicolumn{1}{c}{$P$} &
\multicolumn{1}{c}{$e$} & 
\multicolumn{1}{c}{$f(M)$} &
\multicolumn{1}{c}{T$_0$} &
\multicolumn{1}{c}{$\omega$} & 
\multicolumn{1}{c}{$K_1$} & 
\multicolumn{1}{c}{$V_0$} &
\multicolumn{1}{c}{$a_1 \sin i$} & 
\\
   & &     & 
\multicolumn{1}{c}{(\kms)} &
(d)        &
\multicolumn{1}{c}{(d)} & & 
\multicolumn{1}{c}{(\Msun)}  &
\multicolumn{1}{c}{(JD $-$2\ts400\ts000)} & 
\multicolumn{1}{c}{($^\circ$)} & 
\multicolumn{1}{c}{(\kms)} & 
\multicolumn{1}{c}{(\kms)} & 
($10^6$~km)\\ 
\hline
\smallskip\\
83069 & M2III & 9 & 0.11 & 4450 & $4768\pm175$ & $0.36\pm0.12$ &
0.0076 & $53\ts344\pm106$ & $132\pm15$ & $2.7\pm 0.2$ & $-17.9\pm0.1$
& 163.3 \\
108907 & M4III & 15 & 0.60 & 6821 & $1714\pm4$ & $0.20\pm0.03$ &
0.0094 & $53\ts239\pm40$ & $278\pm9$ & $3.8\pm0.2$&$-15.6\pm0.1$
& 88.6  
& (1)\\
       && 71 & 0.90 & 24830 &  $1703\pm3$ & $0.30\pm0.05$ &
0.0076& $42\ts868\pm40$ & $244\pm9$ & $3.7\pm0.2$&$-14.3\pm0.1$
& 82.0
& (2) \\
       && 72 & 1.02 & 9451 & $1703\pm3$ & $0.33\pm0.02$ &
0.0063& $53\ts204\pm19$ & $267\pm4$ & $3.5\pm0.1$&$-14.8\pm0.1$
& $77.1$
& (3)\\
111129& M2III &  10 & 0.18 & 3633 & $496.7\pm0.7$  & $0.14\pm0.01$ & 
0.047& $50\ts630\pm17$ & $299\pm12$ & $9.8\pm0.3$ & $-15.9\pm0.1$
& 66.3 
\\
113406 & M1III & 12 & 0.25 & 4324 &  $1338\pm7$  & $0.30\pm0.03$ & 
0.0065 & $50\ts865\pm18$ & $18\pm5$ & $3.8\pm0.3$ & $4.8\pm0.1$
& 66.4
\\
137853 & M1III & 11 & 0.12 & 3794 & $1373\pm15$ & $0.15\pm0.06$ &
0.0015 & $51\ts616\pm62$ & $119\pm17$ & $2.2\pm0.1$ & $-9.8\pm0.1$
& 41.7& 
\\
159608&  M2III & 16 & 0.39 & 3981 & $599.3\pm0.4$   & $0.22\pm0.01$& 
0.0751 & $52\ts731\pm5$ & $357\pm3$ & $10.9\pm0.2$ & $-59.9\pm0.1$
& 87.8
\\
165374&  M2III & 37 & 0.41 & 7349 & $2741\pm13$ & $0.05\pm0.02$ &
0.0782& $51\ts058\pm128$ & $74\pm17$ & $6.5\pm0.1$ & $-32.0\pm0.1$
& 245.3  
\\
182190&  M1III & 40 & 0.30 & 5258 & $3856\pm22$& $0.57\pm0.01$ &  0.0155 & $53\ts053\pm11$ & $164\pm2$ & $4.1\pm0.1$ & $-24.4\pm0.1$
& 179.0  \\
187372&  M1III & 16 & 0.36 & 5787 & $986\pm3$     & $0.24\pm0.03$ & 
0.0034& $50\ts201\pm22 $ & $204\pm8$ & $3.3\pm0.2$ & $0.1\pm0.1$
& 43.8  
\\
199871&  M0III & 10 & 0.31 & 3964 & $840.2\pm0.7$   & $0.16\pm0.02$ & 
0.203& $51\ts899\pm7$ & $17\pm3$ & $13.4\pm0.1$ & $-48.2\pm0.1$
& 153.1  
\\
212009&  M0III &  9 & 0.17 & 5008 & $791.8\pm1.1$   & $0.17\pm0.04$ & 
0.0231& $53\ts529\pm15$ & $70\pm6$ & $6.6\pm0.2$ & $-17.7\pm0.2$
&  71.3  
\\
212047&  M4III & 12 & 0.24 & 3984 & $328.2\pm0.1$   & $0.016\pm0.012$&
0.0290& $49\ts315\pm47$ & $176\pm53$ & $9.4\pm0.1$ & $-7.5\pm0.1$
&  42.8  
\\
220088& M0III   & 37 & 0.13 & 5874 & $2516.6\pm1.5$  & $0.639\pm0.004$           & 0.0169 &$53\ts622.5\pm2.6$  & $259.5\pm1.1$ & $5.21\pm0.05$ & $-20.48\pm0.03$ & 138.9 \\
\hline\\
\end{tabular}
}
\end{table*}

\addtocounter{table}{-1}
\begin{table*}
\caption[]{(Continued.)}
{\footnotesize
\begin{tabular}{rlrcrlllllllllllllllll}
\noalign{
\bf Preliminary orbit
}
\smallskip\\
\hline
HD  & Sp. & $N$ &  $\sigma$(O$-$C) & $\Delta t$ &  \multicolumn{1}{c}{$P$} &
\multicolumn{1}{c}{$e$} & Rem.\\
    &     &     &                &   (d)       &  \multicolumn{1}{c}{(d)} & & &\\
\hline\\
203525& M0III &  10 & 0.12 & 4741 & 7290:          & 0.4:          \\
\hline\\
\end{tabular}

\begin{tabular}{r@{}lllllllllllllllllllll}
\noalign{
\bf Orbits from the literature
}
\smallskip\\
\hline
HD  &  & Sp. & $P$  &  $e$ &  $f(M)$    &  Ref. \\
    &  &     & (d)   &      & (\Msun)\\
\hline\\
9053   & & M0III & 193.8 & 0.0 & 0.083 & $\gamma$ Phe: \citet{Luyten-1936}\\	 
41511  & & M4III & $260.34\pm1.80$ & $0.024\pm0.005$ & $0.261\pm0.005$ &  (4)\\
42995  & & M3III & 2983 & 0.53  & 0.13  & $\eta$ Gem: \citet{McLaughlin-1944}\\
80655  & & gM & 834  & 0.0   & 0.0074&  \citet{Griffin-1983}\\  
89758  & & M0III & 230.089$\pm0.039$ & 0.061$\pm0.022$ & 0.01 & $\mu$~UMa:  \citet{Jackson-1957,Lucy-Sweeney-71}\\
108815 & & M     & $4448\pm20$   & $0.73\pm0.03$    & $0.012\pm0.003$ & Griffin, priv. comm.\\ 
111307 & & M0 & $18000\pm3000$ & 0:  & $0.10\pm0.04$  & Griffin, priv. comm.\\
115521 &Aa+b & M2III & $>7000$  & $>0.5$ & .. & Sect.\ref{Sect:115521} \\ 
115521 &A+a &      & $475.1\pm1.4$  & $0.15\pm0.06$ & $9.3\;10^{-5}$  & (5), revised but orbital 
nature doubtful 
(Sect.~\ref{Sect:115521}) \\
117673 & & M0III & 1360.8$\pm2.3$ & 0.26$\pm0.02$  & 0.0095 & \citet{Griffin-2008}\\ 
126947 & & M3III & 2812.3$\pm9.4$   & 0.432$\pm0.021$  & 0.045 &  \citet{Prieur-2006}\\ 
132813 & & M4.5III & 748.9          & 0.13           & 0.0043 & RR UMi (SRb, $P = 43.3$ d):\\
       & &         &                &                &        & \citet{Dettmar-1983,Batten-1986}\\ 
147395 & & M2III & 335.5          & 0.24           & 0.154 & \citet{Carquillat-96}\\
187076 & &  M2II+late B & $3704.2\pm3.2$       & $0.45\pm0.01$           & $0.135\pm0.005$ &  Griffin, priv.comm.\\
       & &            &                       &                       &   & $\delta$~Sge: A $\zeta$~Aur/VV~Cep-like system  
  \citep{Reimers-Schroeder-1983}\\
190658 & & M2.5III & 198.7           & 0.05           & 0.045 &  (6) \\
192867 & & M1III   & 5929           & 0.64           & 0.032 &    \citet{Griffin-1990}\\  
220007 & & M3III   & 1520           & 0.51           & 0.013 &    \citet{Griffin-1979}\\  
\hline\\
\end{tabular}
}
Remarks: (1) this work (2) \citet{Reimers-1988} (3) combined solution (see text)
(4) SS Lep: M4III + accreting A1 star \citep{Welty-1995}. \citet{Verhoelst-2007} shows, from interferometric measurements of the M-giant radius,  that the giant fills its Roche lobe (see also Paper~III) 
(5) Possibly a triple system,
formerly a radial-velocity standard star; 
\citet{Duquennoy-Mayor-91} found a preliminary  509~d period for the inner pair, but the
orbital nature of the short-period variations is questionable (Sect.~\ref{Sect:115521})
(6) HD 190658 = V1472 Aql = HIP 98954: semiregular or most likely
eclipsing or ellipsoidal variable with  $P = 100.37$~d, almost
exactly half the orbital period \citep{Samus-1997}. Orbital elements
from \citet{Lucke-Mayor-1982}. A (physical?) companion is present at
2.3'' \citep{Tokovinin-1997,Frankowski-2007}\\
\end{table*}

\begin{figure}[]
  \includegraphics[width=\columnwidth]{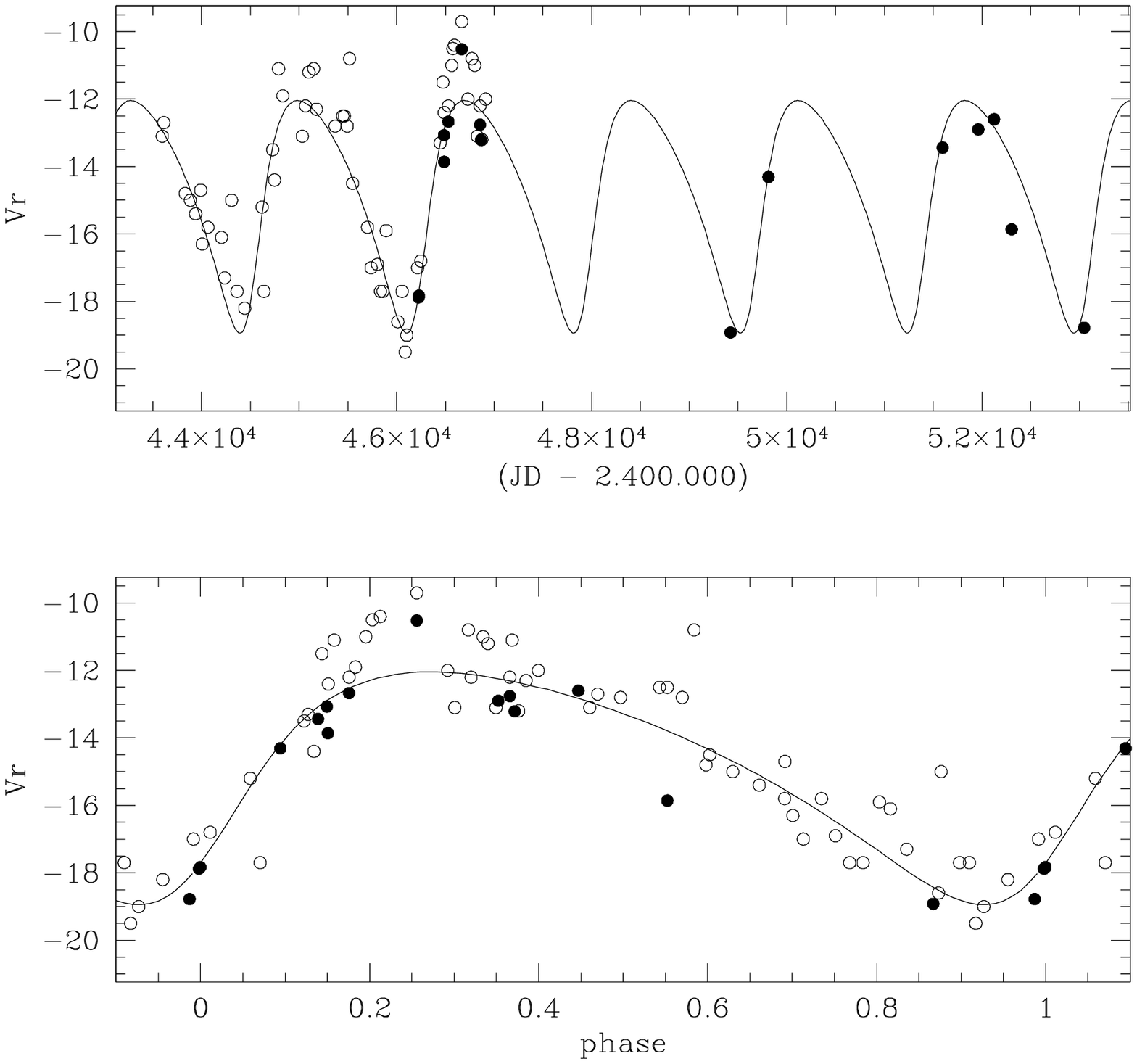}
  \caption{\label{Fig:4Dra}
Top panel: Radial-velocity curve for HD~108907 (4~Dra), an M giant with a hot
compact companion, merging 57 data points from the Cambridge,
Victoria, and CORAVEL spectrovelocimeters listed in
\citet{Reimers-1988} (open symbols) and
the 15 data points from this paper (filled symbols). 
Bottom panel: Phase diagram for the orbital solution merging these two
data sets.
}
\end{figure}

For HD 108907 (4 Dra~=~CQ~Dra), Table~\ref{Tab:orbits} lists three orbital solutions. 
The first entry is obtained with our own CORAVEL and ELODIE data alone. The 
second entry is a solution computed by \citet{Reimers-1988}, based on their 
57 recent data points plus 14 much older ones. The third orbit has been 
computed by merging our 15 data points with the 57 
Cambridge/CORAVEL/Victoria data points from \citet{Reimers-1988}, and
this combined solution is presented in Fig.~\ref{Fig:4Dra}. This
system is of special interest, since 
\citet{Reimers-1985} and \citet{Reimers-1988}  argued
that the companion of the red giant is a cataclysmic
variable, because {\it International Ultraviolet Explorer} spectra
revealed a steep rise shortward of 140~nm along with broad lines (full widths
of 1000~\kms) of highly excited species like He~II and C~IV. 
That conclusion has been challenged, however, by more recent studies 
\citep{Wheatley-2003,Skopal-2005a,Skopal-2005} based on ROSAT X-ray
observations. They conclude instead that the
companion is a single WD
accreting from the wind of its red giant companion, as in normal
symbiotic systems.
The residual radial-velocity jitter of 0.6~\kms\ (listed as $\sigma$(O-C) in Table~\ref{Tab:orbits}) appears normal for a star with
$Sb = 3.8$~\kms, as seen from Fig.~\ref{Fig:Sb-sigma}. 

\acknowledgements{The authors have the pleasure of thanking Roger Griffin for his generous donation of unpublished RV measurements for HD 182190 and HD 220088, which made it possible to compute their orbits. 
We are also indebted for his permission to quote in Table~5 several 
of his new orbits of M giants prior to their publication 
and for his helpful comments on the manuscript of this paper. 
We thank the referee, F. Fekel, whose comments greatly improved the paper, and in particular stimulated the addition of Fig.~3. This work has been partly funded by an {\it Action de recherche concert\'ee (ARC)} from the {\it Direction g\'en\'erale de l'Enseignement non obligatoire et de la Recherche scientifique -- Direction de la recherche scientifique -- Communaut\'e fran\c caise de Belgique.}
}

\bibliographystyle{apj} 
\bibliography{ajorisse_articles} 

\begin{appendix}
\section{Pseudo-orbits among Mira and semiregular variables}

This Appendix presents two cases of (suspected) pseudo-orbital
variations exhibited  by Mira and semi-regular variables. These stars do not belong to the samples studied earlier in this paper.

\subsection{HD~110813}
\label{Sect:110813}
HD~110813 (S~UMa) is a Mira S star with a light cycle of 225.9~d (as listed in the General Catalogue of Variable Stars -- GCVS; the analysis below  shows that a period of 222~d seems more appropriate), and
was considered a spectroscopic binary of period $593\pm62$~d by
\citet{Udry-98a}. It deserves a follow-up discussion here, since recent
datapoints no longer support this orbital solution; in fact, no satisfactory fit can be found to the radial-velocity data. This star provides a good illustration of the difficulties encountered while trying to find spectroscopic binaries among long-period variable stars.

\begin{figure}[]
  \includegraphics[width=\columnwidth]{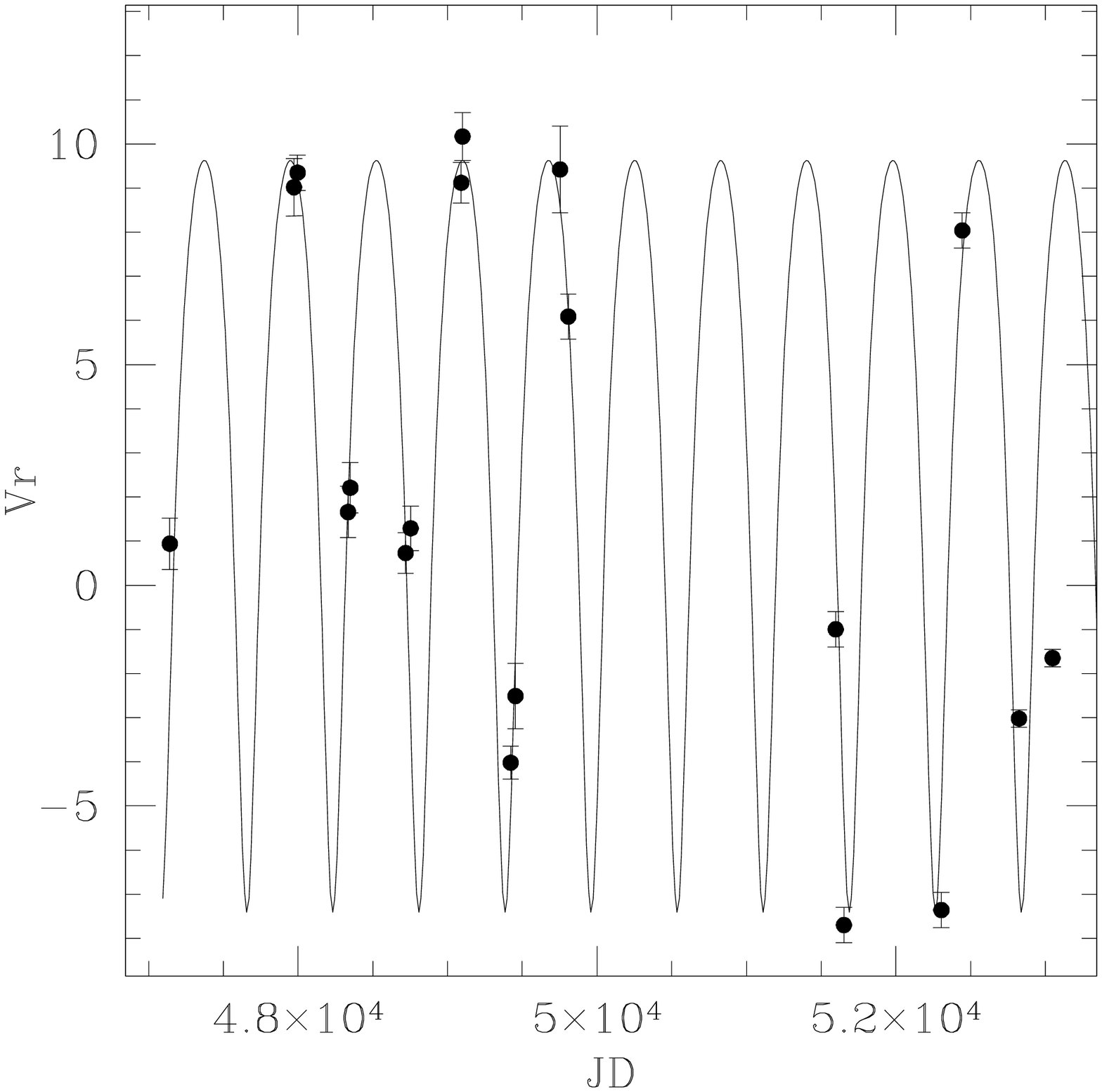}
  \includegraphics[width=\columnwidth]{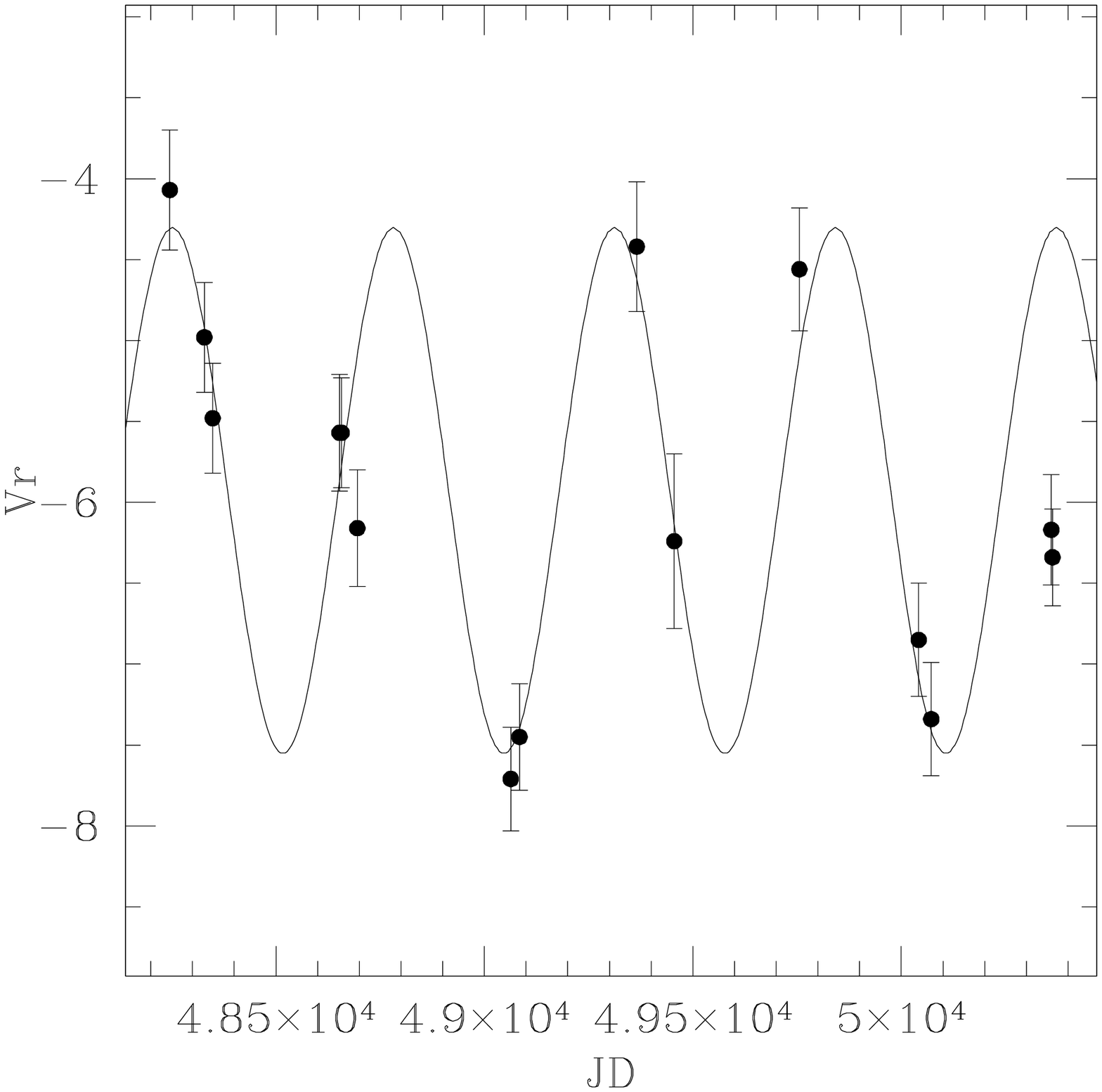}
  \caption{\label{Fig:HD110813} 
Top panel: Radial-velocity curve for HD~110813, a
Mira S star with a pulsation period of 225.9~d (according to GCVS) and a pseudo-orbital
period of 576~d.
Bottom panel: Same for the carbon star HD~76221, a semi-regular
variable with a pulsation period of about 195~d (according to GCVS), 
and a pseudo-orbital period of 530~d. For both stars, the last measurements  become discrepant with the (pseudo-)orbital solution computed from earlier data points. }
\end{figure}

Figure~\ref{Fig:HD110813} shows the radial-velocity curve of S~UMa and
a (pseudo-)orbital solution based on 17 measurements (from 1987.946 to
2002.464 or JD 2\ts 447\ts 141.743 to JD 2\ts 452\ts 444.403;
Tables~\ref{Tab:pulsation} and \ref{Tab:Miradata}). This pseudo-orbit is slightly
different from the one found by \citet{Udry-98a}. Although the
radial-velocity data could be fitted with a period of 576~d for 10 cycles, 
the last
two measurements (JD 2\ts 452\ts 824.356 and JD 2\ts 453\ts 048.562)
deviate markedly from this solution. The radial-velocity variations
cannot therefore be ascribed to orbital motion. In any case, a
system with such a short orbital period cannot be detached. Using the
period--radius relationship for Mira stars pulsating in the fundamental
mode \citep{VanLeeuwen-1997}, a radius of 258~\Rsun\ is inferred from
the 222-d period, assuming a mass of 1.5~\Msun. Adopting a mass of
1~\Msun\ for the companion, for the Roche radius to be larger than the
stellar radius requires an orbital period of at least 1130~d, which is
inconsistent with the observed value of 576~d.

\begin{figure}[]
  \includegraphics[width=\columnwidth]{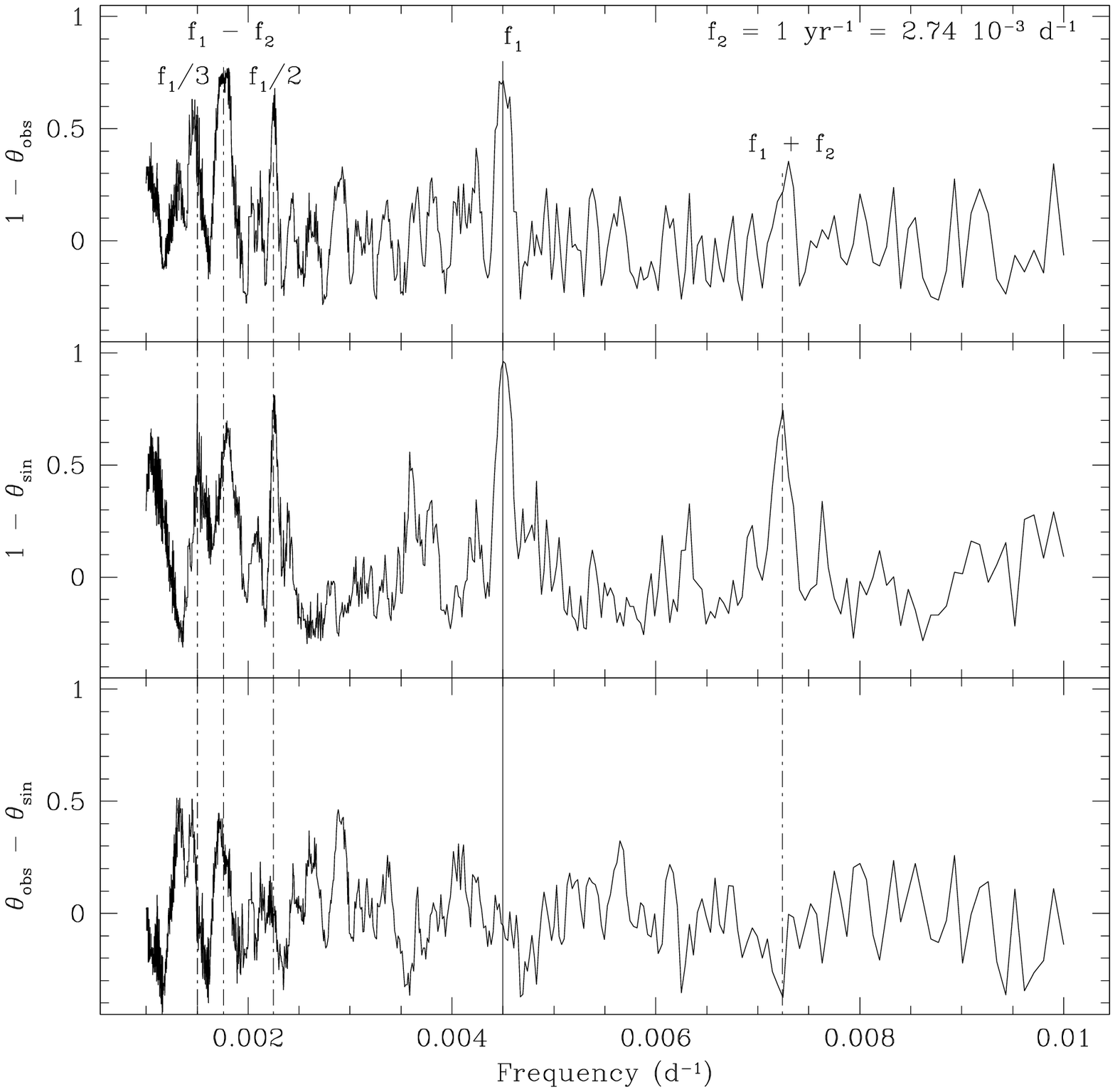}
  \caption{\label{Fig:HD110813theta} 
Top panel: Stellingwerf's
$1-\theta$ statistics (with bins of length 0.2 phase cycle offset by
0.1 phase cycle from the previous one; $N_b = 5$ and $N_c = 2$
adopting Stellingwerf's notations), for all data points of HD~110813
(S~UMa) but the last
two (outliers). Middle panel: the $1-\theta$ statistics for a
sinusoidal signal of period $1/f_1 = 222$~d (corresponding to the Mira
pulsation period) sampled as the data.  Bottom panel: the difference
between the former two, revealing peaks not accounted for by the
aliases associated with the 222-d Mira period.  In the top panel,
several linear combinations of $f_1$ (the Mira frequency) and $f_2$
(the 1-yr frequency) are indicated. The pseudo-orbital period of 576~d
is close to $(f_1 - f_2)^{-1}$ = 555~d.  }
\end{figure}

What then is the origin of this 576~d period?  The Stellingwerf
(phase-dispersion minimisation) $\theta$ statistics
\citep{Stellingwerf-1978} are shown in Fig.~\ref{Fig:HD110813theta},
constructed from all datapoints but the last two outliers. It 
shows that all the prominent peaks are combinations of the Mira frequency $f_1 = 1/222$~d$^{-1}$ and of the yearly frequency $f_2 = 1/365.25$~d$^{-1}$, or harmonics of $f_1$. In particular the 576~d period may be identified with the frequency $f_1 - f_2$.
This finding, along with the previous result for the
minimum period allowed by the Roche radius, definitely denies the
reality of the binary nature of HD~110813.
\citet{Alvarez-2001} have shown that S~UMa exhibits an asymmetric 
cross-correlation dip, as usual among Mira variables. Asymmetric profiles observed in Mira variables are often associated with radial-velocity variations,
which indeed mimick an orbital motion \citep[see also ][]{Hinkle-2002}. 

\subsection{HD~76221}
\label{Sect:76221}

\begin{figure}[]
  \includegraphics[width=\columnwidth]{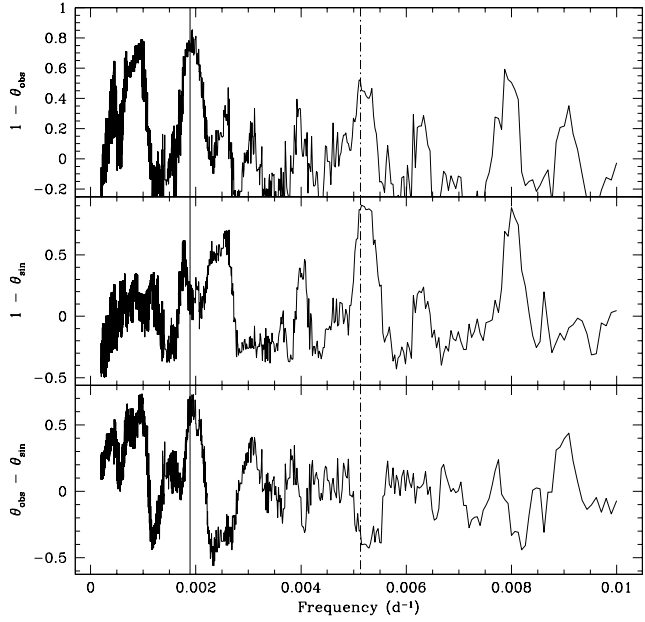}
  \caption{\label{Fig:HD76221theta} 
Top panel: Same as Fig.~\ref{Fig:HD110813theta} for the carbon star
HD~76221, 
for all data points but the last
two (outliers). The solid vertical line marks the frequency of the
pseudo-orbital variation and the dot-dashed line the frequency of
light variations. The harmonics 1/3, 1/2, 3/4, and 3/2 of the pulsation
  frequency 1/195 = $5.1\;10^{-3}$~d$^{-1}$ are cleary identifiable
  in the middle and top panels. Because there is 
a residual peak remaining in the 
lower panel at the pseudo-orbital frequency of 
1/530 = $1.9\;10^{-3}$~d$^{-1}$, this frequency 
does not belong to this series of harmonics. }
\end{figure}

The semiregular carbon star HD~76221 (X~Cnc) behaves similarly to
HD~110813 \citep{Udry-98a}: a satisfactory {(pseudo-)orbital} solution with a period of 530~d (Table~\ref{Tab:pulsation}) could be found
with the first 13 data points, but is not confirmed by the last two data
points (bottom panel of Fig.~\ref{Fig:HD110813}). The phase-dispersion
minimisation statistics (Fig.~\ref{Fig:HD76221theta}, based on the
first 13 data points) reveals several harmonics of the 195~d photometric period,
but this time, the 530-d periodicity of the radial velocities  does not seem to be one of these.   
This 530-d-signal may be yet another example of the long secondary periods
(of unknown origin) found by \citet{Houk-1963} \citep[see also][]{Hinkle-2002} among SR variables, since the 
ranges of periods, mass functions, and 
semi-amplitudes  found by  \citet{Hinkle-2002} (and listed in Table~\ref{Tab:pulsation}) all match the values for
HD~76221. Only the eccentricity does not conform to the Hinkle et
al. range (0.32 -- 0.37, with one case at 0.08). \citet{Soszynski-2007} also notes that the long secondary periods sometimes undergo phase shifts; however, we do not find it very plausible that these photometric and spectroscopic phase shifts have the same physical origin. The phase shifts in the light-curve are attributed by \citet{Soszynski-2007} to dust clouds in the vicinity of a companion, whereas the phase shifts in radial velocity preclude the orbital nature of the radial velocity variations.

\begin{table*}
\caption[]{\label{Tab:pulsation}
Pseudo-orbital solutions for the Mira S star HD~110813 (S~UMa) and the
semi-regular carbon star HD~76221 (X~Cnc), based on the complete data
set but the last two points (see Table~\ref{Tab:Miradata}). 
 }
\begin{tabular}{lllllll}
\hline
        & HD 110813  & HD 76221 & LSP$^a$\\
\hline
$P_{\rm puls}$ (d) & 222 & 195: & \\
'$P_{\rm orb}$' (d) & $576\pm1.9$ & $530\pm 8.9$ & 300 -- 1000\\
$e$     & $0.29\pm0.09$ & 0 & 0.08, 0.32 -- 0.37\\
$\omega$ ($^\circ$) & $178\pm17$ & -- & 240 -- 320\\
$T$ (JD $-$ 2\ts400\ts000) & $52839\pm23$ & $44541\pm75$\\ 
$K$ (\kms) & $8.5\pm0.6$ & $1.6\pm0.2$ & 1.6 -- 3.1\\
$V_0$ (\kms) & $3.6\pm0.5$ & $-5.9\pm0.1$\\
$f(M)$ (\Msun) & $3.2\;10^{-2}$ & $(2.3\pm0.9)\;10^{-4}$ & $2\;10^{-4} - 1.3\;10^{-2}$\\
$a_1 \sin i$ ($10^6$ km) & 64.4 & $11.8\pm1.5$\\
$N$ & 17 & 13\\
$\sigma(O-C)$ (\kms) & 1.82 & 0.44\\
\hline\\
\end{tabular}

$^a$  The ranges found by \citet{Hinkle-2002} for semi-regular
variables  with long secondary periods. 
\end{table*}

\begin{table}
\caption[]{\label{Tab:Miradata}
Radial velocities for HD 76221 and HD~110813.}
\begin{tabular}{r@{.}l@{}r@{.}lccr}
\hline
\multicolumn{2}{c}{JD $-$ 2\ts 400\ts 000} & \multicolumn{2}{c}{$Vr$} & \multicolumn{1}{c}{$\epsilon$} & \multicolumn{1}{c}{$O-C$} & Instr.\\
\multicolumn{2}{c}{} & \multicolumn{2}{c}{(\kms)} & \multicolumn{1}{c}{(\kms)} & \multicolumn{1}{c}{(\kms)}\\
\hline
\multicolumn{4}{c}{}
\medskip\\
\multicolumn{4}{c}{}\\
\noalign{\hspace*{\fill}\bf HD 76221\hspace*{\fill}}
\medskip\\
48245&634 &  $-$4&07 &0.37 &$+$0.20 &COR \\
48328&471 &  $-$4&98 &0.34 &$-$0.05&COR  \\
48348&421 &  $-$5&48 &0.34 &$-$0.21&COR  \\
48652&522 &  $-$5&57 &0.36 &$+$0.33 &COR \\
48657&563 &  $-$5&57 &0.34 &$+$0.23 &COR \\
48695&451 &  $-$6&16 &0.36 &$-$1.07&COR  \\
49063&382 &  $-$7&71 &0.32 &$-$0.21&COR  \\
49084&380 &  $-$7&45 &0.33 &$-$0.07&COR  \\
49365&587 &  $-$4&42 &0.40 &$+$0.19 &COR \\
49455&376 &  $-$6&24 &0.54 &$-$0.06&COR  \\  
49755&526 &  $-$4&56 &0.38 &$+$0.52 &COR \\
50041&685 &  $-$6&85 &0.35 &$+$0.25 &COR \\
50071&650 &  $-$7&34 &0.35 &$+$0.06 &COR \\
50359&680 &  $-$6&17 &0.34 & --   &COR\\
50362&648 &  $-$6&34 &0.33 & --   &COR\\
\multicolumn{4}{c}{}
\medskip\\
\multicolumn{4}{c}{}\\ 
\noalign{\hspace*{\fill}\bf HD 110813\hspace*{\fill}}
\medskip\\
47141&743  &  0&94& 0.58 & $+$3.38 & COR \\
47972&573  &  9&02& 0.65 & $-$0.49 & COR  \\
47996&476  &  9&35& 0.40 & $+$0.14 & COR \\
48334&584  &  1&66& 0.58 & $-$0.41 & COR \\
48349&556  &  2&21& 0.57 & $-$1.22 & COR \\
48719&460  &  0&73& 0.46 & $-$0.24 & COR \\
48753&455  &  1&29& 0.50 & $+$4.35 & COR \\
49091&467  &  9&12& 0.46 & $-$0.48& COR \\
49100&422  & 10&17& 0.54 & $+$0.56 & COR \\
49422&565  & $-$4&02& 0.37 & $+$1.13 & COR \\
49454&561  & $-$2&51& 0.74 & $-$1.09& COR \\
49753&604  &  9&42& 0.98 & $+$0.87 & COR \\
49808&417  &  6&09& 0.51 & $-$0.12 & COR \\
51597&5    & $-$1&0 & 0.40 & $-$2.22 & ELO\\
51652&5    & $-$7&7 & 0.40 & $-$2.46 & ELO\\
52304&611  &$-$7&36 &0.40 & $-$2.39 & ELO\\
52444&403  &   8&04 &0.40 & $+$0.68 & ELO\\
52824&356  &$-$3&02 &0.40 & $-$ & ELO\\      
53048&562  &$-$1&65 &0.40  & $-$ & ELO\\
\hline
\end{tabular}
\end{table}

\end{appendix}

\end{document}